\documentclass[apj]{emulateapj}
\usepackage{graphics}
\usepackage{subfigure}
\usepackage[caption=false]{subfig}
\usepackage{epsfig}
\usepackage{alltt}
\usepackage{verbatim}

\usepackage[usenames,dvipsnames]{xcolor}
\definecolor{royalblue}{HTML}{4169E1}
\definecolor{green}{HTML}{008000}

\def\spose#1{\hbox to 0pt{#1\hss}}
\def\simlt{\mathrel{\spose{\lower 3pt\hbox{$\mathchar"218$}}
    \raise 2.0pt\hbox{$\mathchar"13C$}}}
\def\simgt{\mathrel{\spose{\lower 3pt\hbox{$\mathchar"218$}}
    \raise 2.0pt\hbox{$\mathchar"13E$}}}

\newcommand{\ch}{{\it Chandra} }
\newcommand{\oiiin}{\mbox{[\ion{O}{3}]}}
\newcommand{\oiii}{\mbox{[\ion{O}{3}]} $\,$}
\newcommand{\oiiiw}{\mbox{[\ion{O}{3}] $\lambda$5007} $\,$}
\newcommand{\oiiiwn}{\mbox{[\ion{O}{3}] $\lambda$5007}}

\newcommand{\hb}{\mbox{H$\beta$} $\,$}
\newcommand{\hbn}{\mbox{H$\beta$}}
\newcommand{\ha}{\mbox{H$\alpha$} $\,$}
\newcommand{\han}{\mbox{H$\alpha$}}
\newcommand{\oiiihb}{\mbox{[\ion{O}{3}] $\lambda$5007}/{\mbox{H$\beta$} $\,$}}

\newcommand{\niihan}{\mbox{[\ion{N}{2}] $\lambda$6584}/{\mbox{H$\alpha$}}}

\newcommand{\siihan}{\mbox{[\ion{S}{2}] $\lambda\lambda$6717,6731}/{\mbox{H$\alpha$}}}

\newcommand{\justniin}{\mbox{[\ion{N}{2}]}}
\newcommand{\justsii}{\mbox{[\ion{S}{2}]} $\,$}
\newcommand{\justsiin}{\mbox{[\ion{S}{2}]}}
\newcommand{\justoin}{\mbox{[\ion{O}{1}]}}

\newcommand{\oi}{\mbox{[\ion{O}{1}] $\lambda$6300} $\,$}
\newcommand{\oihan}{\mbox{[\ion{O}{1}] $\lambda$6300}/{\mbox{H$\alpha$}}}

\newcommand{\cl}{\mbox{[\ion{Ne}{5}] $\lambda$3347}, \mbox{[\ion{Ne}{5}] $\lambda$3427}, \mbox{[\ion{Fe}{7}] $\lambda$3586}, \mbox{[\ion{Fe}{7}] $\lambda$3760}, or \mbox{[\ion{Fe}{7}] $\lambda$6086}}
\newcommand{\hi}{\mbox{\ion{H}{1}} $\,$} 

\slugcomment{Accepted for publication in ApJ}
\shortauthors{Comerford et al.}
\shorttitle{Towards a More Complete Optical Census of Active Galactic Nuclei}

\begin{document}

\title{Towards a More Complete Optical Census of Active Galactic Nuclei, \\ Via Spatially-Resolved Spectroscopy}

\author{Julia M. Comerford\altaffilmark{1}, James Negus\altaffilmark{1}, R. Scott Barrows\altaffilmark{1}, Dominika Wylezalek\altaffilmark{2}, Jenny E. Greene\altaffilmark{3}, Francisco M\"{u}ller-S\'{a}nchez\altaffilmark{4}, and Rebecca Nevin\altaffilmark{5}}

\affil{$^1$Department of Astrophysical and Planetary Sciences, University of Colorado, Boulder, CO 80309, USA}

\affil{$^2$Zentrum f\"{u}r Astronomie der Universit\"{a}t Heidelberg Astronomisches Rechen-Institut M\"{o}nchhofstr, 12-14 69120 Heidelberg, Germany}

\affil{$^3$Department of Astrophysical Sciences, Princeton University, Princeton, NJ 08544, USA}

\affil{$^4$Department of Physics and Materials Science, University of Memphis, 3720 Alumni Avenue, Memphis, TN 38152, USA}

\affil{$^5$Center for Astrophysics $\|$ Harvard \& Smithsonian, 60 Garden St., Cambridge, MA 02138, USA}

\begin{abstract}

While emission-line flux ratio diagnostics are the most common technique for identifying active galactic nuclei (AGNs) in optical spectra, applying this approach to single fiber spectra of galaxies can omit entire subpopulations of AGNs.  Here, we use spatially resolved spectroscopy from the  Mapping Nearby Galaxies at Apache Point Observatory (MaNGA) survey to construct a sample of 10 galaxies where Baldwin-Philips-Terlevich line flux ratio diagnostics classify each galaxy's central $3^{\prime\prime}$  spectrum as LINER or star forming, while $>10\%$ of the spaxels in the galaxy's MaNGA footprint are classified as Seyfert. We obtain {\it Chandra} observations of these 10 galaxies with off-nuclear Seyfert regions to determine whether AGNs are actually present in them.  Our main result is that 7-10 (depending on strictness of criteria) of the galaxies host one or more X-ray AGNs, even though none of them were classified as AGNs based on their single-fiber optical spectra.  We find that these AGNs were not identified in the single-fiber spectra because they are AGNs in the nuclei of companion galaxies, low luminosity AGNs, dust obscured AGNs, and/or flickering AGNs. In summary, we find that off-nuclear AGN signatures may increase the number of known AGNs by a factor of two over what conventional single nuclear fiber spectra identify.  Our results show that spatially resolved spectroscopy can be leveraged to reveal a more complete census of AGNs that are traditionally missed by single fiber spectra.

\end{abstract}  

\keywords{AGN host galaxies, Active galactic nuclei, X-ray active galactic nuclei, Low-luminosity active galactic nuclei, Galaxies, LINER galaxies, Seyfert galaxies, Star formation}

\section{Introduction}
\label{intro}

Over the past 20 years, the Sloan Digital Sky Survey (SDSS) has revolutionized our understanding of active galactic nuclei (AGNs) and galaxy evolution. This success was built, in part, on the thousands of nearby AGNs identified through the $3^{\prime\prime}$ optical single fiber spectra used in the SDSS-I through SDSS-III surveys. While some of these AGNs (the Type 1 AGNs) were identified by broad emission lines, the majority of AGNs in SDSS were identified using the Baldwin-Philips-Terlevich (BPT; \citealt{BA81.1, VE87.1}) emission line ratio diagrams, which classify the source of the ionizing radiation as star formation (SF), a low-ionization nuclear emission-line region (LINER; \citealt{HE80.1}), Seyfert, or composite combination of Seyfert and star formation.  Varieties of the BPT diagram include the \justniin-BPT diagram (\oiiihb vs. \niihan) and the \justsiin-BPT diagram (\oiiihb vs. \siihan).  The \justniin-BPT diagram distinguishes between star formation, Seyferts, and composite, while the \justsiin-BPT diagram distinguishes between star formation, Seyferts, and LINERs.

However, many studies have illustrated that these BPT diagnostics can systematically misclassify the sources of emission in a galaxy, since the emission line ratios themselves can be changed by effects including radiation from post-asymptotic giant branch stars (post-AGB stars) and shocks (e.g., \citealt{RI11.1,KE13.1,BE16.1}).  Further, there is no single source of LINER emission. Depending on the spatial scale and other environmental factors, LINERs may be caused by weak AGNs (e.g., \citealt{ST97.1,HO99.1,BA02.3}), photoionization from sources of hard radiation (e.g., \citealt{FE83.1,BI94.1,YA12.1}), and shock ionization (e.g., \citealt{HE80.1,DO95.1,AL08.1}).

BPT diagnostics are more ambiguous when they are based on a single spectrum of a galaxy, such as a $3^{\prime\prime}$ SDSS fiber spectrum.  Spatially resolved spectroscopy can reveal some of the nuances of BPT classifications.  For example, spatially resolved spectroscopy shows that LINER emission is not always nuclear; many galaxies have spatially-extended, non-nuclear, LINER emission (e.g., \citealt{PH86.1,HO14.2,BE16.1}).  As another example, spatially resolved spectroscopy has shown that star formation is responsible for much of the systematic offset between local galaxies and $z>1$ galaxies on BPT diagrams (e.g., \citealt{KE13.1,HI21.1}).

Using a single $3^{\prime\prime}$ optical fiber spectrum, which covers only the central 5 kpc of a $z=0.1$ SDSS galaxy, has other limitations in identifying AGNs.  This approach can miss obscured AGNs, AGNs in the nuclei of companion galaxies (e.g., \citealt{GR11.1,LI13.1,BA16.1,CO17.1}), and flickering AGNs that leave limited emission (sometimes only hard emission) in the central regions but extended light echoes out at larger galactocentric distances (e.g., \citealt{LI09.2,KE15.1,CO17.2}).  In many of these cases, the galaxies can exhibit off-nuclear optical AGN emission even though the galaxy center has little or no optical signature of an AGN.  Spatially resolved spectroscopy again can help detect such systems.  For instance, \cite{ME20.1} used spatially resolved spectroscopy of a sample of dwarf galaxies with off-nuclear AGN emission to determine that the off-nuclear emission is caused by off-nuclear AGNs or AGNs that have turned off and left behind an echo of past ionization.

Here, we use spatially resolved spectroscopy from the SDSS-IV survey Mapping Nearby Galaxies at Apache Point Observatory (MaNGA; \citealt{BU15.1,LA15.1}) to explore a set of 10 galaxies at $z<0.15$ where the central $3^{\prime\prime}$ spectrum indicates SF or LINER, yet $>10\%$ of the spaxels in the MaNGA footprint are classified as Seyfert \citep{WY18.1}.  We obtain {\it Chandra X-ray Observatory} observations of these galaxies to determine whether the off-nuclear Seyfert regions are signatures that the galaxies host AGNs, and we analyze the {\it Chandra} and MaNGA observations in tandem to determine why these galaxies have such unexpected spatially-resolved BPT maps.  The results of this analysis showcase how spatially resolved spectroscopy can drive a more complete census of AGNs.

The rest of this paper is organized as follows.  In Section~\ref{sample} we describe how we selected the 10 MaNGA galaxies for {\it Chandra} observations.  Section~\ref{obs} outlines our analysis of the {\it Chandra} and MaNGA observations.  Section~\ref{emission} explains how we use the {\it Chandra} observations to identify AGNs, as well as how we use the emission lines observed by MaNGA to identify the sources of gas ionization.  In Section~\ref{results}, we present our results for the galaxy population as a whole, and Section~\ref{nature} contains our interpretation of each individual galaxy.  Finally, Section~\ref{conclusions} presents our conclusions.

We assume a Hubble constant $H_0 =70$ km s$^{-1}$ Mpc$^{-1}$, $\Omega_m=0.3$, and $\Omega_\Lambda=0.7$ throughout, and all distances are given in physical (not comoving) units.

\section{The Galaxy Sample}
\label{sample}

We build our sample from the galaxies observed by MaNGA, which is an SDSS-IV integral field spectroscopy (IFS) survey of $10,010$ low-redshift galaxies.  MaNGA obtained its data from 2014 to 2020 \citep{BU15.1,DR15.1,LA15.1,YA16.1,BL17.1,WA17.1}.  MaNGA uses IFS with $2^{\prime\prime}$ fibers grouped into hexagonal bundles, which range in diameter from $12\farcs5$ to $32\farcs5$.  The PSF FWHM is $2\farcs5$.  The observations span 3600 - 10,300 \AA \, with a spectral resolving power of $R\sim2000$, and the redshift range is $0.01<z<0.15$ (average redshift $z=0.03$).  MaNGA targets galaxies with stellar masses $>10^9$ $M_\odot$, and the survey was designed to spectroscopically map galaxies out to at least 1.5 times the effective radius.

Our parent sample consists of the 2727 galaxies that have been observed in the fifth MaNGA Product Launch (MPL-5).  From this parent sample, \cite{WY18.1} used BPT emission line diagnostics, as well as cuts on H$\alpha$ surface brightness and equivalent width, to identify AGN candidates in these MaNGA galaxies.  The AGN candidates were identified as the systems that met the following four criteria: 

1) The fraction of spaxels that are classified as `AGN' or `Composite' in the \justniin-BPT diagram is greater than 10\%, and the mean value of the top 20\% percentile of the distribution of \ha equivalent widths in spaxels that are classified as `AGN' or `Composite' in the \justniin-BPT diagram is greater than 5 \AA.

 2) The fraction of spaxels that are classified as `AGN' or `LINER' in the \justsiin-BPT diagram is greater than 15\%, and the mean value of the top 20\% percentile of the distribution of \ha equivalent widths in spaxels that are classified as `AGN' or `LINER' in the \justsiin-BPT diagram is greater than 5 \AA.
 
 3) $\log$ (SB(\han)$_{A,N}$/(erg s$^{-1}$ kpc$^{-2}$)) $> 37.5$, where SB(\han)$_{A,N}$ is the mean \ha surface brightness of the spaxels that are classified as `AGN' or `Composite' in the \justniin-BPT diagram; or $\log$ (SB(\han)$_{AL,S}$/(erg s$^{-1}$ kpc$^{-2}$)) $> 37.5$, where SB(\han)$_{AL,S}$ is the mean \ha surface brightness of the spaxels that are classified as `AGN' or `LINER' in the \justsiin-BPT diagram.
 
 4) For the 20\% of the spaxels that are classified as `AGN' or `LINER' in the \justsiin-BPT diagram that have the largest distances from the star formation demarcation line, the mean distance is greater than 0.3.  Each spaxel's distance is defined such that the line connecting the spaxel measurement in the \justsiin-BPT diagram and the star formation demarcation line is minimized.

These criteria, which have been optimized for MaNGA, identify 303 AGN candidates.  Interestingly, the majority of these AGN candidates (173 out of 303) would not have been selected as AGNs based on the central single-fiber spectrum.  Rather, the \justsiin-BPT diagram classifies the single-fiber, central $3^{\prime\prime}$ spectrum as star-forming or LINER  (from the Portsmouth catalog of \citealt{TH13.1}).   We use the \justsiin-BPT diagnostics because they enable classification of LINER sources.  Here, we define the ``off-nuclear Seyfert region" sample of galaxies as those AGN candidates where the central $3^{\prime\prime}$ fiber spectrum is classified as star-forming or LINER but $>10\%$ of the spaxels in the MaNGA footprint are classified as Seyfert. 

Our aim is to use {\it Chandra} observations to determine whether these galaxies in fact host AGNs.  For the follow-up {\it Chandra} observations, we chose the off-nuclear Seyfert region galaxies where the Seyfert spaxels have a summed \oiiiw flux that is $>5 \times 10^{-15}$ erg cm$^{-2}$ s$^{-1}$, since large \oiiiw fluxes minimize the exposure times needed with {\it Chandra} (Section~\ref{chandra}).  This yielded eight targets: four with star-forming central regions and four with LINER central regions.  We also cross-matched the off-nuclear Seyfert region sample with {\it Chandra} archival data, and we found two additional galaxies that have archival {\it Chandra} observations: one with a star-forming central region and one with a LINER central regions.  Our complete sample is made up of these 10 galaxies (Table~\ref{tbl-1}).

\begin{deluxetable*}{llllll}
\tablewidth{0pt}
\tablecolumns{6}
\tablecaption{Measurements from SDSS Observations} 
\tablehead{
\colhead{SDSS Designation} &
\colhead{$3^{\prime\prime}$ fiber classif.} &
\colhead{$z$} & 
\colhead{$\log M_* (\log M_\odot$)} & 
\colhead{$\log$ SFR ($\log M_\odot \, \mathrm{yr}^{-1}$)} & 
\colhead{$E(B-V)^a$}
}
\startdata 
SDSS J074351.36+444327.5 & SF & $0.031$  & $10.5 \pm 0.1$ & $0.01 \pm 0.06$ & $0.56 \pm 0.01$ \\     
SDSS J074507.25+460420.6 & LINER & $0.031$ & $11.0 \pm 0.1$ & $0.33 \pm 0.06$ & $0.41 \pm 0.07$ \\
SDSS J093106.75+490447.1 &  SF & $0.034$ & $11.0 \pm 0.1$ & $0.91 \pm 0.06$ & $0.63 \pm 0.01$ \\
SDSS J125448.95+440920.1 &   LINER & $0.054$ &  $11.2 \pm 0.1$ & $0.64 \pm 0.06$ & $0.28 \pm 0.05$ \\
SDSS J140737.17+442856.2 W &  LINER & $0.143$ & $9.6$ & $-0.43 \pm 0.02$ & $0.59 \pm 0.22$  \\
SDSS J140737.17+442856.2 E &  &  & $9.4$ & $0.27 \pm 0.04$  & $0.26 \pm 0.05$ \\
SDSS J143031.19+524225.8  & SF & $0.045$ & $11.2 \pm 0.1$ & $1.07 \pm 0.06$  & $0.78 \pm 0.01$ \\
SDSS J151806.13+424445.0 NW & SF & $0.040$ & $9.2$ & $0.42 \pm 0.03$ & $0.81 \pm 0.01$  \\
SDSS J151806.13+424445.0 SE &  & & $9.3$ & $-0.40 \pm 0.08$ & $0.60 \pm 0.01$  \\     
SDSS J160153.01+452107.0 & LINER & $0.042$ & $11.0 \pm 0.1$ & $0.22 \pm 0.06$  & $0.12 \pm 0.01$ \\   
SDSS J163014.63+261223.3 & SF & $0.131$  & $11.3 \pm 0.1$ & $1.23 \pm 0.06$ & $0.44 \pm 0.01$  \\         
SDSS J163342.33+391106.5 & LINER & $0.030$ & $10.5 \pm 0.1$ & $-0.27 \pm 0.06$  & $0.37 \pm 0.03$ 
\enddata
\tablecomments{$^a$For the spaxel centered on the stellar bulge.}
\label{tbl-1}
\end{deluxetable*}

\section{Observations and Analysis}
\label{obs}

\begin{deluxetable*}{lll}
\tablewidth{0pt}
\tablecolumns{3}
\tablecaption{Summary of {\it Chandra} Observations} 
\tablehead{
\colhead{SDSS Name} &
\colhead{{\it Chandra}/ACIS} & 
\colhead{{\it Chandra}/ACIS} \\
\colhead{} &
\colhead{Exp. Time (s)} & 
\colhead{Obs. Date (UT)}
}
\startdata 
SDSS J0743+4443 & 16884 / 15885 & 2018 Dec 30 / 2019 Jan 04 \\  
SDSS J0745+4604 & 33654 & 2018 Dec 24 \\
SDSS J0931+4904 & 29699 & 2019 Jan 12 \\
SDSS J1254+4409 & 34637 & 2019 Oct 08 \\
SDSS J1407+4428$^a$ & 29684 & 2017 Feb 25 \\
SDSS J1430+5242 & 23841 & 2020 Mar 21 \\
SDSS J1518+4244$^a$ & 14470 & 2006 Sep 11 \\
SDSS J1601+4521 & 19835 & 2019 Dec 24 \\
SDSS J1630+2612 & 29704 & 2018 Dec 11 \\
SDSS J1633+3911 & 23778 & 2019 Nov 14
\enddata
\tablecomments{$^a$Archival {\it Chandra} observations.}
\label{tbl-2}
\end{deluxetable*}

\begin{deluxetable*}{lllll} 
\tablewidth{0pt}
\tablecolumns{5}
\tablecaption{{\it Chandra} Source Properties in Different Energy Ranges} 
\tablehead{
\colhead{SDSS Name} &
\colhead{{\it Chandra} Restframe} & 
\colhead{Counts} &
\colhead{$L_{X,abs}$} &
\colhead{$L_{X,unabs}$}  \\
\colhead{} &
\colhead{Energy Range (keV)} &
\colhead{} &
\colhead{($10^{40}$ erg s$^{-1}$)} &
\colhead{($10^{40}$ erg s$^{-1}$)} 
}
\startdata 
SDSS J0743+4443 & $0.5-2$ & $8.8^{+2.4}_{-3.7}$ & $0.3^{+0.2}_{-0.6}$ & $0.6^{+0.4}_{-0.7}$ \\
 & $2-8$ & $4.6^{+1.6}_{-2.9}$ & $0.4^{+0.3}_{-0.9}$ & $0.9^{+0.6}_{-1.0}$ \\
\hline
SDSS J0745+4604 & $0.5-2$ & $93.5^{+9.0}_{-10.4}$ & $4.2^{+1.0}_{-1.4}$ & $9.7^{+2.1}_{-2.1}$ \\
 & $2-8$ & $93.8^{+9.1}_{-10.4}$ & $8.7^{+2.2}_{-2.8}$ & $20.1^{+4.3}_{-4.3}$ \\
\hline
SDSS J0931+4904 & $0.5-2$ & $32.2^{+5.1}_{-6.3}$ & $2.4^{+0.7}_{-0.9}$ & $3.1^{+0.9}_{-1.0}$ \\
 & $2-8$ & $22.5^{+4.0}_{-5.4}$ & $3.0^{+0.9}_{-1.1}$ & $3.9^{+1.1}_{-1.2}$ \\
\hline
SDSS J1254+4409 & $0.5-2$ & $8.8^{+2.4}_{-3.7}$ & $<3.7$ & $<500$ \\ 
 & $2-8$ & $23.2^{+4.4}_{-5.5}$ & $14.6^{+7.0}_{-9.5}$ & $147.6^{+67.3}_{-66.1}$ \\
\hline
SDSS J1407+4428 E & $0.5-2$ & $108.1^{+9.7}_{-11.0}$ & $239.3^{+51.0}_{-56.4}$ & $320.3^{+65.6}_{-66.9}$ \\
& $2-8$ & $930.5^{+30.7}_{-30.4}$ & $2538.9^{+541.3}_{-598.5}$ & $3398.3^{+696.1}_{-709.3}$ \\
\hline
SDSS J1430+5242 & $0.5-2$ & $1036.0^{+29.8}_{-34.4}$ & $158.9^{+18.5}_{-19.6}$ & $216.6^{+21.1}_{-23.9}$ \\
 & $2-8$ & $1747.4^{+41.1}_{-42.7}$ & $575.0^{+67.0}_{-70.8}$ & $783.9^{+76.2}_{-86.5}$ \\
\hline
SDSS J1518+4244 NW & $0.5-2$ & $70.8^{+8.4}_{-9.8}$ & $5.3^{+2.6}_{-3.9}$ & $10.9^{+4.9}_{-4.9}$ \\ 
 & $2-8$ & $23.5^{+4.6}_{-5.8}$ & $4.4^{+2.2}_{-3.3}$ & $9.1^{+4.1}_{-4.1}$ \\
SDSS J1518+4244 SE & $0.5-2$ & $70.8^{+8.4}_{-9.8}$ & $0.9^{+0.3}_{-0.3}$ & $0.9^{+0.3}_{-0.3}$ \\ 
 & $2-8$ & $23.5^{+4.6}_{-5.8}$ & $1.0^{+0.4}_{-0.4}$ & $1.1^{+0.4}_{-0.4}$ \\
\hline
SDSS J1601+4521 & $0.5-2$ & $16.8^{+3.5}_{-4.7}$ & $3.4^{+1.0}_{-0.9}$ & $4.0^{+1.2}_{-1.0}$ \\ 
 & $2-8$ & $22.2^{+4.3}_{-5.4}$ & $4.9^{+1.4}_{-1.2}$ & $5.7^{+1.7}_{-1.4}$ \\
\hline
SDSS J1630+2612 & $0.5-2$ & $16.5^{+3.8}_{-4.6}$ & $9.3^{+3.9}_{-5.7}$ & $25.7^{+10.}_{-12.2}$ \\
 & $2-8$ & $540.9^{+22.3}_{-24.4}$ & $1563.5^{+657.1}_{-963.8}$ & $4318.7^{+1677.7}_{-2048.1}$ \\
\hline
SDSS J1633+3911 & $0.5-2$ & $12.5^{+2.8}_{-4.2}$ & $2.3^{+1.7}_{-8.4}$ & $4.9^{+3.4}_{-9.4}$ \\
 & $2-8$ & $6.8^{+2.0}_{-3.3}$ & $0.7^{+0.5}_{-2.5}$ & $1.5^{+1.0}_{-2.8}$ 
\enddata
\label{tbl-3}
\end{deluxetable*}

\subsection{Chandra/ACIS X-ray Observations and Analysis}
\label{chandra}

Eight of the off-nuclear Seyfert region galaxies were observed with {\it Chandra}/ACIS for the program GO9-20089X (PI: Comerford).  We derived exposure times from the summed \oiiiw flux of the Seyfert spaxels in each system and the scaling relation between \oiiiw flux and hard X-ray (2-10 keV) flux for Type 2 AGNs, which has a scatter of 1.06 dex \citep{HE05.1}.  We selected exposure times that would ensure a firm detection of at least $3\sigma$ of each AGN and measurements of the extragalactic column density that are accurate to an order of magnitude or better.  The galaxies were observed with exposure times of 20 ks to 35 ks (Table~\ref{tbl-2}).  The observations of SDSS J0743+4443 were split across two observing dates, and we registered each observation to SDSS broadband imaging (see below) and then used {\tt merge\_obs} to merge these two exposures.  The merged observation was used for the image modeling described below.

For the remaining two off-nuclear Seyfert region galaxies in our sample, we analyzed archival {\it Chandra} observations.  SDSS J1407+4428 (ObsID = 19990; PI: Secrest) was observed for 30 ks, while SDSS J1518+4244 (ObsID =6858, PI: Komossa) was observed for 14 ks.

The galaxies were observed with the telescope aimpoint on the ACIS S3 chip in ``timed exposure'' mode and telemetered to the ground in ``faint'' mode.  We reduced the data with the latest {\it Chandra} software (CIAO\,4.13) in combination with the most recent set of calibration files (CALDB\,4.9.4).  

First, we registered the \ch observations following the same approach as the MaNGA registration \citep{LA16.2}, so that we can accurately compare the positions of the X-ray and optical sources.   We registered against the SDSS broadband imaging in the $g$, $r$, $i$, and $z$ bands, and used a biweight mean of the four bands as our final result.

For each galaxy, we used {\tt dmcopy} to make a sky image of the field in the rest-frame soft (0.5-2 keV), hard (2-8 keV), and total (0.5-8 keV) energy ranges. Then, with the modeling facilities in {\tt Sherpa}, we simultaneously modeled each X-ray source as a 2D Lorenztian function ({\tt beta2d}: $f (r) = A(1 + [r / r_0]^2) - \alpha$) and the background as a fixed count rate estimated from an annulus of 2\farcs5 width around the source region. We used the {\tt wavdetect} source position that is closest to the galaxy centroid as the initial input position for each {\tt beta2d} fit.  Then, we allowed the model to fit a region of radius $2\farcs5$, which is more than double the {\it Chandra} point spread function (PSF) radius. To determine the best-fit model parameters, we used {\tt Sherpa}'s implementation of the ``Simplex" minimization algorithm \citep{LA98.1} and minimized the Cash statistic.

Our sample includes two merging galaxy systems with two stellar bulges each, SDSS J1407+4428 and SDSS J1518+4244, and we attempted a two-component {\tt beta2d} model to test for additional X-ray sources in these systems.  We set the initial positions of the two {\tt beta2d} components to the stellar bulge positions and allowed them to wander within a circle of radius $2\farcs5$.  In SDSS J1518+4244 we detected two X-ray sources with significances $>3\sigma$ above the background, and these two sources are spatially coincident with the two stellar bulges.  The $2\farcs5$ radius extraction regions did not overlap for these two sources.  For SDSS J1407+4428, we detected an X-ray source at the position of the eastern nucleus with a significance $32\sigma$ above the background. We found a western source with a significance of $2.6\sigma$ above the background, and since this is a $<3\sigma$ result we do not classify it as a detection.  In total, we find 11 X-ray sources with significances $>3\sigma$ above the background: the eight nonmerging galaxies each have an X-ray source coincident with their centers, SDSS J1407+4428 has an X-ray source coincident with the eastern nucleus, and SDSS J1518+4244 has two X-ray sources (one coincident with each of the two stellar bulges).

Next, we used the Bayesian Estimation of Hardness Ratios ({\tt BEHR}) code \citep{PA06.1} to measure the rest-frame soft, hard, and total counts in each X-ray source. We used {\tt calc\_data\_sum} to determine the number of observed soft and hard counts from both the source region and a background region, and then {\tt BEHR} uses a Bayesian approach to estimate the expected values and uncertainties of the rest-frame soft counts and rest-frame hard counts. Table~\ref{tbl-3} shows these values.

We then used {\tt Sherpa} to model the energy spectra of the extracted regions over the observed energy range 0.5-7 keV.  We fit each unbinned spectrum with a redshifted power law, $F \sim E^{-\Gamma}$, which represents the intrinsic AGN X-ray emission at the galaxy redshift.  This spectrum is attenuated by passing through two absorbing column densities of neutral Hydrogen. One of these is fixed to the Galactic value, which we determined using an all-sky interpolation of the \hi in the Galaxy \citep{DI90.1}, and the other ($n_H$) is assumed to be intrinsic to the source at the galaxy redshift.

For the fit to each spectrum, we allowed $\Gamma$ and $n_H$ to vary freely.  For three sources, we found that the best-fit value of $\Gamma$ was not within the typical range of observed power-law indices, i.e. $1\le \Gamma \le 3$ \citep{NA94.2,RE00.1,PI05.1,IS10.1}, and for these we redid the fits with $\Gamma$ fixed at a value of 1.8, which is a typical value for the continuum of Seyfert galaxies.  

To determine the best-fit model parameters for each spectrum, we used \texttt{Sherpa}'s implementation of the Levenberg-Marquardt optimization method \citep{BE69.1} to minimize the Cash statistic.  For SDSS J0743+4443 the observations were split across two observing dates, and so we extracted a spectrum from each observation and modeled them simultaneously.  Table~\ref{tbl-4} shows the results of these spectral fits.  For the three sources where we redid the fits with fixed $\Gamma=1.8$, we use the results of the $\Gamma=1.8$ fits for the analyses that follow.  

All fluxes are $k$-corrected, and we calculated the observed flux values from the model sum (including the absorbing components) and the intrinsic flux values from the unabsorbed power law component.  Then, we used the redshift to determine the distance to each system and convert the X-ray fluxes to X-ray luminosities (Table~\ref{tbl-3}).  Finally, we converted the rest-frame 2-10 keV luminosities to bolometric luminosities by multiplying by a factor of 20, which is a typical bolometric correction for AGNs (e.g., \citealt{EL94.1,MA04.4}).

\begin{deluxetable}{llll} 
\tablewidth{0pt}
\tablecolumns{4}
\tablecaption{{\it Chandra} Spectral Fits} 
\tablehead{
\colhead{SDSS Name} &
\colhead{$n_H$} & 
\colhead{$\Gamma$} &
\colhead{Reduced} \\
\colhead{} &
\colhead{($10^{21}$ cm$^{-2}$)} &
\colhead{} &
\colhead{C-stat} 
}
\startdata 
SDSS J0743+4443 & $<$3E-6 & $2.3^{+0.6}_{-0.8}$ & $0.09$ \\
SDSS J0745+4604 & $<$6E-7 & $2.5^{+0.1}_{-0.2}$ & $0.53$ \\
SDSS J0931+4904 & $<$7E-7 & $2.1^{+0.2}_{-0.3}$ & $0.28$ \\
SDSS J1254+4409 & $10.8^{+4.5}_{-4.0}$ & $9.0^{+2.6}_{-1.6}$ & $0.30$ \\ 
 & $365.4^{+120.8}_{-103.1}$ & $1.8^a$ & $0.25$ \\
SDSS J1407+4428 E & $10.4^{+2.2}_{-2.1}$ & $1.1^{+0.1}_{-0.1}$ & $0.69$ \\ 
SDSS J1430+5242 & $3.0^{+0.6}_{-0.6}$ & $1.6^{+0.1}_{-0.1}$ & $0.76$ \\
SDSS J1518+4244 NW & $3.6^{+1.6}_{-1.6}$ & $2.8^{+0.5}_{-0.5}$ & $0.30$ \\ 
SDSS J1518+4244 SE & $3.9^{+2.5}_{-2.7}$ & $3.8^{+0.9}_{-1.0}$ & $0.15$ \\ 
 & $0.0^{+0.5}_{-0.6}$ & $1.8^a$ & $0.16$ \\
SDSS J1601+4521 & $<$4E-7 & $0.5^{+0.4}_{-0.5}$ & $0.25$ \\ 
 & $<$6E-7 & $1.8^a$ & $0.25$ \\
SDSS J1630+2612 & $89.8^{+6.8}_{-10.9}$ & $1.5^{+0.1}_{-0.2}$ & $0.68$ \\
SDSS J1633+3911 & $<$1E-6 & $3.0^{+1.6}_{-1.7}$ & $0.14$  
\enddata
\tablecomments{$^a$The best-fit spectrum had a $\Gamma$ outside of the usual AGN range $1 \leq \Gamma \leq 3$, so we redid the fit by freezing $\Gamma=1.8$.  For these three sources, we use the results of the $\Gamma=1.8$ spectral fits in the rest of our analyses.}
\label{tbl-4}
\end{deluxetable}

\begin{deluxetable*}{llllll}  
\tablewidth{0pt}
\tablecolumns{6}
\tablecaption{{\it Chandra} Evidence for AGN} 
\tablehead{
\colhead{SDSS Name} &
\colhead{$L_{2-10 \mathrm{keV}, \mathrm{XRB}}$} &
\colhead{$L_{2-10 \mathrm{keV}, \mathrm{unabs}}$} &
\colhead{$L_{2-10 \mathrm{keV}, \mathrm{XRB}}$ /} &
\colhead{Sigma} &
\colhead{{\it Chandra} evidence}  \\
\colhead{} &
\colhead{($10^{40}$ erg s$^{-1}$)} &
\colhead{($10^{40}$ erg s$^{-1}$)} &
\colhead{$L_{2-10 \mathrm{keV}, \mathrm{unabs}}$} &
\colhead{significance} &
\colhead{for AGN?}
}
\startdata 
SDSS J0743+4443 & $0.4 \pm 0.2$ & $1.1^{+0.7}_{-1.3}$ & 0.39 & 4 & yes$^a$ \\ 
SDSS J0745+4604 & $1.2 \pm 0.2$ & $24.4^{+5.2}_{-5.2}$ & 0.05 & $>100$ & yes \\ 
SDSS J0931+4904 & $2.1 \pm 0.3$ & $5.0^{+1.4}_{-1.6}$ & 0.43  & 11 & yes$^a$ \\ 
SDSS J1254+4409  & $2.3 \pm 0.2$ & $293.1^{+133.6}_{-131.3}$ & $<0.01$ & $>100$ & yes \\ 
SDSS J1407+4428 E & $0.3 \pm 0.2$ & $5478.5^{+1122.0}_{-1143.9}$ & $<0.01$ & $>100$ & yes \\ 
SDSS J1430+5242  & $3.2 \pm 0.3$ & $1100.9^{+107.1}_{-121.5}$ & $<0.01$ & $>100$ & yes \\ 
SDSS J1518+4244 NW  & $0.4 \pm 0.2$ & $10.5^{+4.7}_{-4.7}$ & 0.04 & 55 & yes \\
SDSS J1518+4244 SE & $0.1 \pm 0.2$ & $1.5^{+0.5}_{-0.6}$ & 0.06 & 9 & yes \\ 
SDSS J1601+4521  & $1.2 \pm 0.2$ & $7.6^{+2.2}_{-1.9}$ & 0.15 & 33 & yes \\ 
SDSS J1630+2612  & $4.7 \pm 0.4$ & $7080.3^{+2750.7}_{-3358.1}$ & $<0.01$ & $>100$ & yes \\ 
SDSS J1633+3911  & $0.4 \pm 0.2$ & $1.7^{+1.2}_{-3.2}$ & 0.23 & 7 & yes$^a$  
\enddata
\tablecomments{Column 2 shows the estimated restframe 2-10 keV luminosity from XRBs (Section~\ref{agn}).  Column 3 shows the restframe, unabsorbed 2-10 keV luminosity that we measure from the {\it Chandra} observations (Section~\ref{chandra}).  Column 4 is the ratio of the estimated restframe 2-10 keV luminosity from XRBs to the restframe, unabsorbed 2-10 keV luminosity.  Column 5 indicates the sigma significance with which $L_{2-10 \mathrm{keV}, \mathrm{unabs}}$ is greater than $L_{2-10 \mathrm{keV}, \mathrm{XRB}}$.  Column 6 indicates whether an AGN is present: if $L_{2-10 \mathrm{keV}, \mathrm{unabs}}$ is more than $3\sigma$ (Column 5) greater than $L_{2-10 \mathrm{keV}, \mathrm{XRB}}$, then this is evidence for an AGN. $^a$These galaxies have weaker AGN detections, since the XRBs contribute more than 20\% to the total unabsorbed 2-10 keV luminosity and the significance of the X-ray luminosity in excess of the expected contribution from XRBs is $\lesssim 10\sigma$.}
\label{tbl-5}
\end{deluxetable*}

\subsection{MaNGA Analysis}
\label{manga}

Our analyses of the MaNGA galaxies rely on galaxy properties measured in the Pipe3D Value Added Catalog \citep{SA16.1,SA18.2}, including galaxy stellar mass and star formation rate (SFR; derived from \ha measurements).  For the two merging galaxy systems, we used the stellar mass of each bulge as measured from best-fit models of the nuclear spectra in \cite{FU18.1}.  For these two merging galaxy systems, we measured a separate SFR for a $2\farcs5$ diameter aperture centered on each stellar bulge.  We summed the \ha flux in each aperture and converted to SFR via the relation between \ha luminosity and SFR given in \cite{KE12.1}, which is the same approach used to measure the SFRs in Pipe3D that we use for the remaining eight galaxies.   These values are shown in Table~\ref{tbl-1}, and we note that they may be overestimates of the SFR if an AGN is present and contributing to the \ha luminosity. 

We also used the emission line fluxes measured in MaNGA's Data Analysis Pipeline (DAP; \citealt{WE19.1}).  We use the non-parametric summed fluxes of the emission lines, where these fluxes are measured after subtraction of the stellar continuum model, and Galactic reddening is also accounted for.  Further, we use the DAP's non-parametric equivalent width measurements of \han.  

Finally, we also use the MaNGA emission line fluxes to measure the Balmer decrement, $\han/\hbn$, which provides a measurement of the dust attenuation \citep{CA00.1,YU18.1}.  We then use the Balmer decrement to determine the color excess $E(B-V)$ \citep{YU18.1}, and in Table~\ref{tbl-1} we present our measurements of $E(B-V)$ in the spaxel corresponding to the central $2\farcs5$ ($\sim1.5$-6 kpc) of each galaxy's stellar bulge(s). 

\section{Nature of the Emission}
\label{emission}

\subsection{X-ray AGN Identifications} 
\label{agn}

To determine whether the {\it Chandra} observations indicate the presence of an X-ray AGN, we first account for the X-ray binary (XRB) contribution to the hard X-rays.  For each galaxy we calculate the predicted hard X-ray luminosity from XRBs $L_{2-10 \mathrm{keV}, \mathrm{XRB}}$, as a function of galaxy stellar mass and SFR, using the relations in \cite{LE10.1}.  We use the galaxy stellar masses and SFRs described in Section~\ref{manga}, which includes separate stellar mass and SFR measurements for each bulge in the merging galaxy systems.  We then determine the uncertainties on $L_{2-10 \mathrm{keV}, \mathrm{XRB}}$ from propagating the scatter (0.34 dex) in the \cite{LE10.1} relation, the coefficient uncertainties in the \cite{LE10.1} relation, the uncertainties on galaxy stellar mass, and the uncertainties on SFR.  We define an X-ray AGN detection as those cases where $L_{2-10 \mathrm{keV}, \mathrm{unabs}}$ is more than $3\sigma$ greater than $L_{2-10 \mathrm{keV}, \mathrm{XRB}}$ (e.g., \citealt{BA19.2}).  By this definition, all 11 X-ray detections presented in this paper are X-ray AGNs (Table~\ref{tbl-5}).

However, for three galaxies (SDSS J0743+4443, SDSS J0931+4904, and SDSS J1633+3911), the XRBs contribute more than 20\% to the total unabsorbed 2-10 keV luminosity and the significance of the X-ray luminosity in excess of the expected contribution from XRBs is $\lesssim 10\sigma$.  To be conservative, we label these three sources as weaker AGN detections (Table~\ref{tbl-5}).  Therefore, $7-10$ ($70-100\%$) of the galaxies in our sample have at least one confirmed X-ray AGN, depending on how conservative our criteria are.

\begin{deluxetable*}{lllllllll}
\footnotesize
\tablewidth{0pt}
\tablecolumns{9}
\tablecaption{AGN Classification by Method} 
\tablehead{
\colhead{SDSS Name} &
\colhead{BPT, WHAN} &
\colhead{BPT, \ha EW} &
\colhead{Coronal} &
\colhead{Radio} &
\colhead{{\it WISE}} & 
\colhead{Broad Line} &
\colhead{{\it Swift}/} &
\colhead{{\it Chandra}} \\
\colhead{} &
\colhead{of $3^{\prime\prime}$ MaNGA} &
\colhead{of $3^{\prime\prime}$ SDSS} &
\colhead{Lines in} &
\colhead{(HERG,} &
\colhead{Color} &
\colhead{in $3^{\prime\prime}$ SDSS} &
\colhead{BAT} &
\colhead{(This Paper)} \\
\colhead{} &
\colhead{Spectrum$^a$} &
\colhead{Spectrum$^b$} &
\colhead{MaNGA$^c$} &
\colhead{LERG)} &
\colhead{} &
\colhead{Spectrum} &
\colhead{} &
\colhead{} 
}
\startdata 
SDSS J0743+4443   & no & no & no & no & no & no  & no & \textcolor{green}{yes$^d$} \\  
SDSS J0745+4604   & \textcolor{green}{yes} & \textcolor{green}{yes} & no & no & no & no  & no & \textcolor{green}{yes}  \\
SDSS J0931+4904   & no & no & no & no & \textcolor{green}{yes} & no  & no & \textcolor{green}{yes$^d$}  \\
SDSS J1254+4409   & \textcolor{green}{yes} & \textcolor{green}{yes} & no & no & no & no  & no  & \textcolor{green}{yes} \\
SDSS J1407+4428  & \textcolor{green}{yes} & no & no & no & \textcolor{green}{yes} & no  & no & \textcolor{green}{yes}   \\
SDSS J1430+5242   & no & no & no & no & no & no  & no & \textcolor{green}{yes}  \\
SDSS J1518+4244   & no & no & no & no & \textcolor{green}{yes} & no  & no & \textcolor{green}{yes}  \\
SDSS J1601+4521  & \textcolor{green}{yes} & \textcolor{green}{yes} & no & no & no & no  & no & \textcolor{green}{yes}   \\
SDSS J1630+2612   & no & \textcolor{green}{yes} & no & no & \textcolor{green}{yes} & no  & no & \textcolor{green}{yes}  \\
SDSS J1633+3911 & \textcolor{green}{yes} & \textcolor{green}{yes} & no & no & no & no  & no  & \textcolor{green}{yes$^d$}
\enddata
\tablecomments{The sample was selected so that the galaxies are AGN candidates in \cite{WY18.1}, but not AGN according to the \justsiin-BPT diagram of the central $3^{\prime\prime}$ SDSS spectrum. Columns 2 and 3 illustrate approaches to selecting AGN {\it candidates} via SDSS and MaNGA spectra ($^a$\citealt{SA18.2}; $^b$\citealt{RE17.1}). Column 4 shows AGNs selected via coronal emission lines in the MaNGA observations ($^c$\citealt{NE21.1}).  Columns 5-8 show AGNs {\it confirmed} via multiwavelength observations in the MaNGA AGN catalog \citep{CO20.1}. Column 9 shows that all 10 galaxies host AGNs confirmed via the {\it Chandra} observations analyzed in this paper.  $^d$These galaxies have weaker {\it Chandra} AGN detections, since the XRBs contribute more than 20\% to the total unabsorbed 2-10 keV luminosity and the significance of the X-ray luminosity in excess of the expected contribution from XRBs is $\lesssim 10\sigma$ (see Table~\ref{tbl-5}).} 
\label{tbl-6}
\end{deluxetable*} 

\subsection{Sources of Ionization}
\label{ionization}

While BPT diagrams mainly focus on SF and AGNs as the ionization sources producing emission lines, here we broaden the analysis of these MaNGA galaxies by also considering ionization by post-AGB stars and shocks.   Post-AGB stars are capable of producing hard ionized spectra (e.g., \citealt{BI94.1,YA12.1}), but they are not thought to produce \ha equivalent widths greater than $\sim3$ \AA \, (e.g., \citealt{BE16.1}).  Consequently, a cutoff of  \ha equivalent width $<3$ \AA \, can identify regions of a galaxy that are predominantly ionized by post-AGB stars \citep{CI10.3}.  We present such individual spaxel by spaxel maps in Figure~\ref{fig:maps}.

The \oi emission line is an indicator of shocks (e.g., \citealt{DO76.1,AL08.1,RI11.1}), and line flux ratios of \oihan $>0.1$ indicate that shocks with velocities 160-300 km s$^{-1}$ are the main excitation source of \oi  \citep{RI21.1}. Further, the line flux ratio cutoff \siihan $>0.4$ has also been established empirically to be an identifier of supernova remnant (SNR) shocks \citep{DO78.1, DO80.1}.   Figure~\ref{fig:maps} shows individual spaxel by spaxel maps of these emission line ratio cutoffs for the MaNGA galaxies in our sample, and we use them for the analyses of the individual galaxies that follow (Section~\ref{nature}).

\section{Results}
\label{results}

We have found that $70-100\%$ of the off-nuclear Seyfert region galaxies indeed host AGNs, as shown by the {\it Chandra} observations (Section~\ref{agn}).  Here, we explore whether there are any unusual properties of this sample that could explain why we found X-ray AGNs in this sample even though the central $3^{\prime\prime}$ fiber spectrum of each galaxy was not classified as Seyfert emission.

\subsection{Two of the AGNs Are Not Detected Via Other Approaches: Evidence of AGN Flickering} 
\label{other}

We selected our galaxies from the \cite{WY18.1} sample of AGN candidates in MaNGA, and we found that all of them have AGNs based on {\it Chandra} observations.  Now, we explore whether these systems have AGN detections in other wavelength regimes and using other approaches.  Table~\ref{tbl-6} summarizes our results.

First, we crossmatch our sample to the MaNGA AGN catalog \citep{CO20.1}, which is a catalog of MaNGA galaxies where an AGN was detected via mid-infrared {\it WISE} colors (using colors cuts from \citealt{AS18.1}), {\it Swift}/BAT ultra hard X-ray detections (using the 105-month BAT catalog of AGNs; \citealt{OH18.1}), NRAO Very Large Array Sky Survey (NVSS) and Faint Images of the Radio Sky at Twenty centimeters (FIRST) radio observations (using the catalog of radio sources corresponding to AGNs; \citealt{BE12.1}), and broad emission lines detected in single-fiber SDSS spectra \citep{OH15.1}.  The radio sources are further subdivided into high-excitation radio galaxies (HERGs) and low-excitation radio galaxies (LERGs).  Each dataset in the MaNGA AGN catalog covers the full sample of MaNGA galaxies considered here.  While we find that four of the galaxies in our sample have AGNs identified by {\it WISE} colors, none of them are identified as AGN by {\it Swift}/BAT, radio HERG or LERG classifications, or broad emission lines (which would only identify Type I AGNs).

Next, we crossmatch our sample to MaNGA catalogs of AGN candidates that were selected spectroscopically.  \cite{SA18.2} selected AGN candidates based on the integrated spectra of the central $3^{\prime\prime}$ by $3^{\prime\prime}$ of MaNGA galaxies.  To classify an object as an AGN candidate, they require that this central integrated spectrum have emission line ratios that lie above the theoretical maximum for starbursts in the BPT diagram \citep{KE01.2, KE06.1} and an \ha equivalent width that is $> 1.5 \, \AA \,$ \citep{CI10.3}.  Out of 2755 galaxies in MaNGA MPL-5, they identify 98 AGN candidates.  We find that five of our galaxies are AGN candidates based on these criteria.

In addition, \cite{RE17.1} searched for AGNs in MaNGA using both BPT and WHAN (where \ha equivalent widths $>3$ \AA \, identify ionization by AGNs, and not ionization by post-AGB stars; \citealt{CI10.3}) diagnostics simultaneously.  Using the SDSS-III integrated nuclear spectra of the 2778 galaxies observed in MaNGA MPL-5, they identify 62 `true' AGNs whose line flux ratios and \ha equivalent widths lie in the Seyfert or LINER regions of both the BPT and WHAN diagrams.  Five of our galaxies are included in this sample of AGNs.

We also crossmatch our sample to a study of coronal line emitting galaxies in MaNGA, where coronal lines are emission lines with high ionization potentials ($\simgt100$ eV) that are suggestive of AGN activity.  \cite{NE21.1} use the full MaNGA footprint of each galaxy to search for emission from one or more coronal lines (\cl) detected at $\geq 5\sigma$ above the background continuum in at least 10 spaxels.  Using these criteria, they find 10 coronal line emitting galaxies out of 6263 galaxies observed in MaNGA MPL-8.  None of these 10 are galaxies in our sample.

Finally, we note that two galaxies (SDSS J0743+4443 and SDSS J1430+5242) host AGNs that are identified by {\it Chandra} but not by any other approach, and in Section~\ref{nature} we suggest that these are AGNs are flickering, where the X-rays are the signature of current AGN activity.

\begin{deluxetable}{lll}
\tablewidth{0pt}
\tablecolumns{3}
\tablecaption{AGN Bolometric Luminosity Comparison} 
\tablehead{
\colhead{SDSS Name} &
\colhead{{\it WISE} log $L_{bol}$} &
\colhead{{\it Chandra} log $L_{bol}$}  
}
\startdata 
SDSS J0931+4904 & $44.3 \pm 43.4$ & $42.0 \pm 41.5$ \\
SDSS J1407+4428 & $45.1 \pm 44.2$ & $45.0 \pm 44.4$ \\
SDSS J1518+4244 & $44.5 \pm 43.6$ & $42.4 \pm 42.0^a$ \\ 
SDSS J1630+2612 & $45.3 \pm 44.3$ & $45.2 \pm 44.8$
\enddata
\tablecomments{For the galaxies with AGNs detected in both {\it WISE} and {\it Chandra}. $^a$The NW and SE X-ray luminosities are summed.}
\label{tbl-7}
\end{deluxetable}

\begin{figure*}
\begin{center}
\includegraphics[width=7.in]{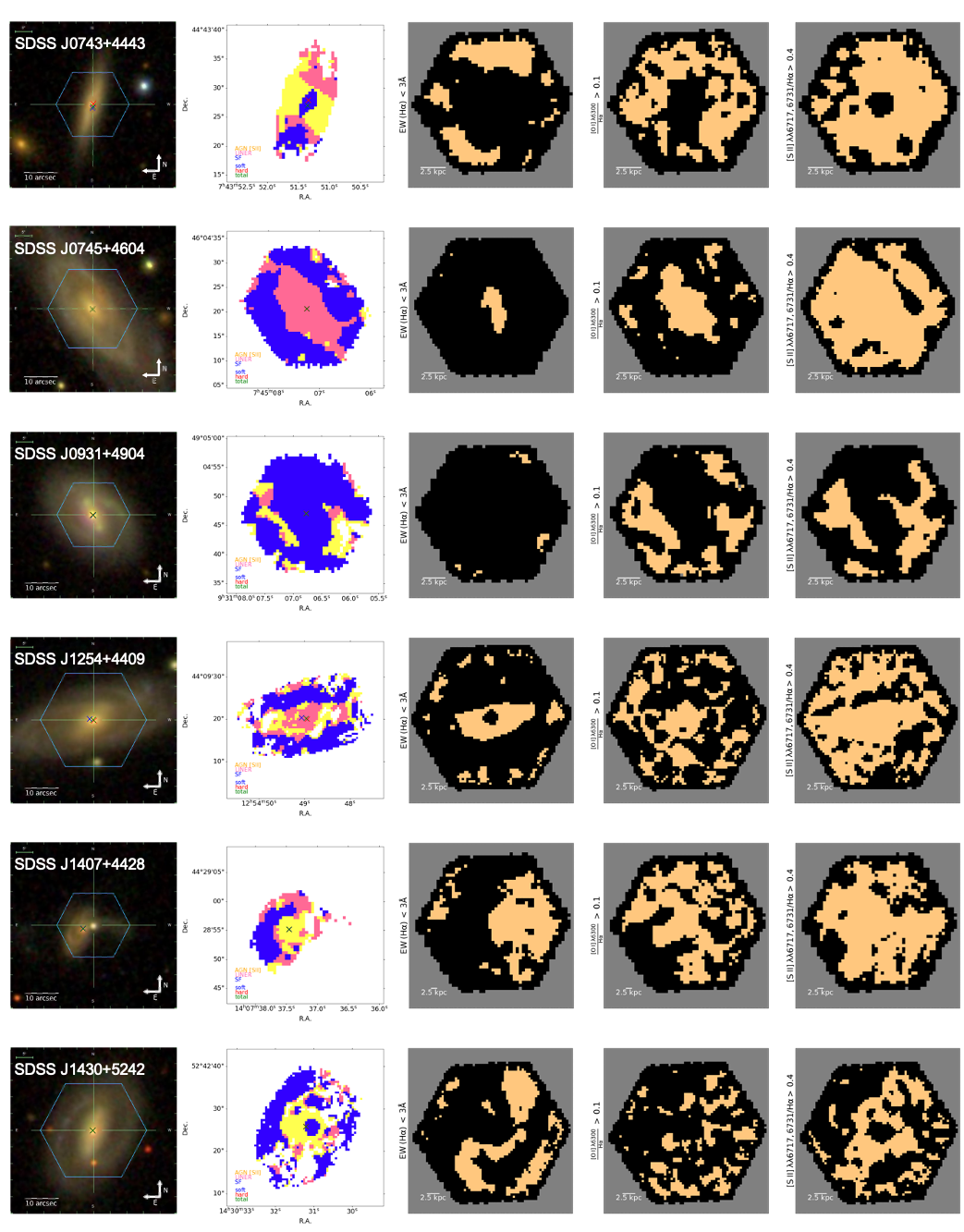}
\end{center}
\end{figure*}

\begin{figure*}
\centering
\includegraphics[width=7.in]{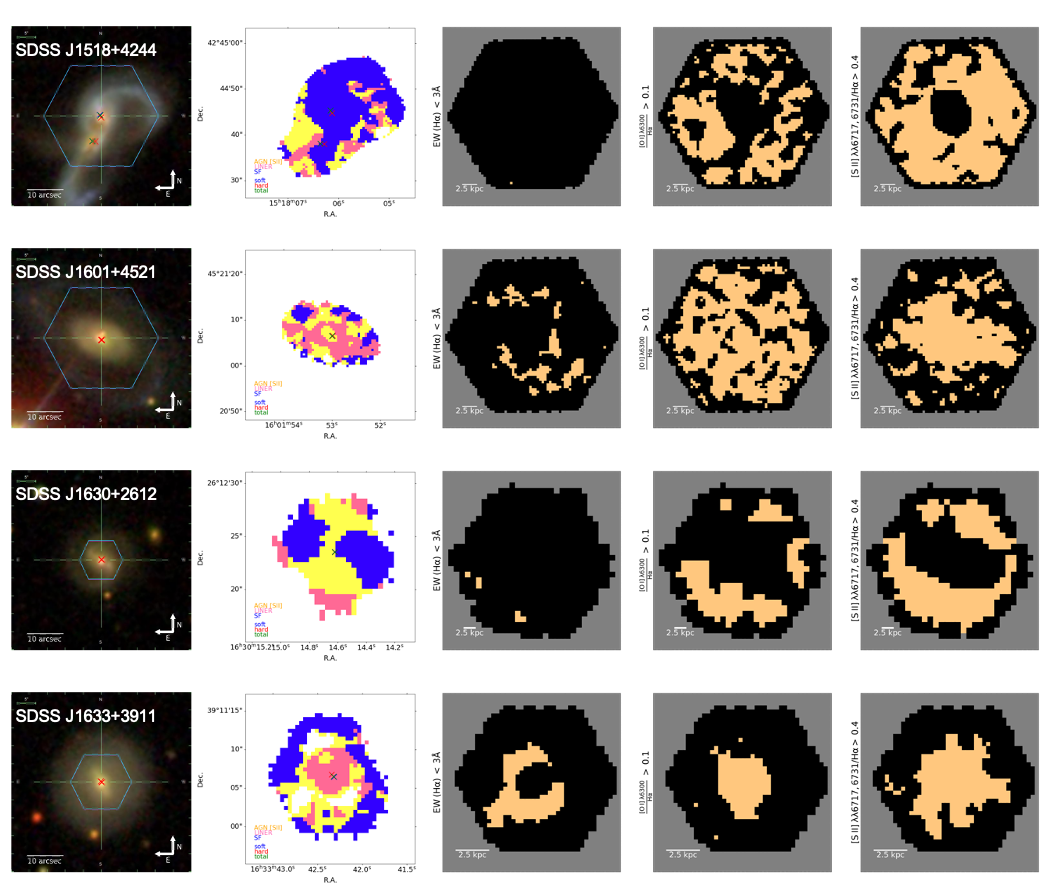}
\caption{Left to right: SDSS image of the galaxy with MaNGA footprint shown as the blue hexagon; \justsiin-BPT diagram, where yellow indicates Seyfert, pink indicates LINER, and blue indicates SF; regions of the galaxy where the equivalent width of \ha is $<3$ \AA, indicating that the emission is dominated by post-AGB stars, are shown in yellow; regions of the galaxy where \oihan $>0.1$, indicating that shocks are present, are shown in yellow; regions of the galaxy where \siihan $>0.4$, indicating that SNR shocks are present, are shown in yellow.  In the leftmost two panels, the blue, red, and green crosses illustrate the positions of the soft, hard, and total X-rays, respectively.}  
\label{fig:maps}
\end{figure*}

\subsection{Two Galaxies Have High $L_{X,2-10\mathrm{keV}}/L_{\oiiin}$: Evidence of AGN Flickering}
\label{LxLOIII}

Since the X-ray and optical classifications of these 10 galaxies are different, we compare the hard X-ray luminosity and optical emission line luminosity (in this case, \oiiiw luminosity) to explore whether this galaxy sample is unusual in its X-ray or optical luminosities.   We use the \oiiiw observed fluxes of the central $3^{\prime\prime}$ spectra from SDSS DR7, as measured by the OSSY catalog \citep{OH11.1}, and the absorbed 2-10 keV luminosities measured in Section~\ref{chandra}.

The fiducial relation between these two luminosities for Type 2 AGNs is log($L_{X,2-10\mathrm{keV}}/L_{\oiiiwn})=0.57 \pm 1.06$, based on a sample of 29 Type 2 AGNs at $z<0.2$ \citep{HE05.1}.  However, Type 2 dual AGNs fall systematically below this relation, suggesting that mergers drive excess gas onto the AGNs that increase their \oiiiw luminosities, and/or that mergers induce higher nuclear gas columns that suppress the hard X-ray luminosities of the AGNs \citep{LI13.1, CO15.1, BA17.2}.  This trend for dual AGNs illustrates that the location of an AGN on this plot can reveal unusual circumstances.

\begin{figure}
\begin{center}
\includegraphics[width=3.in]{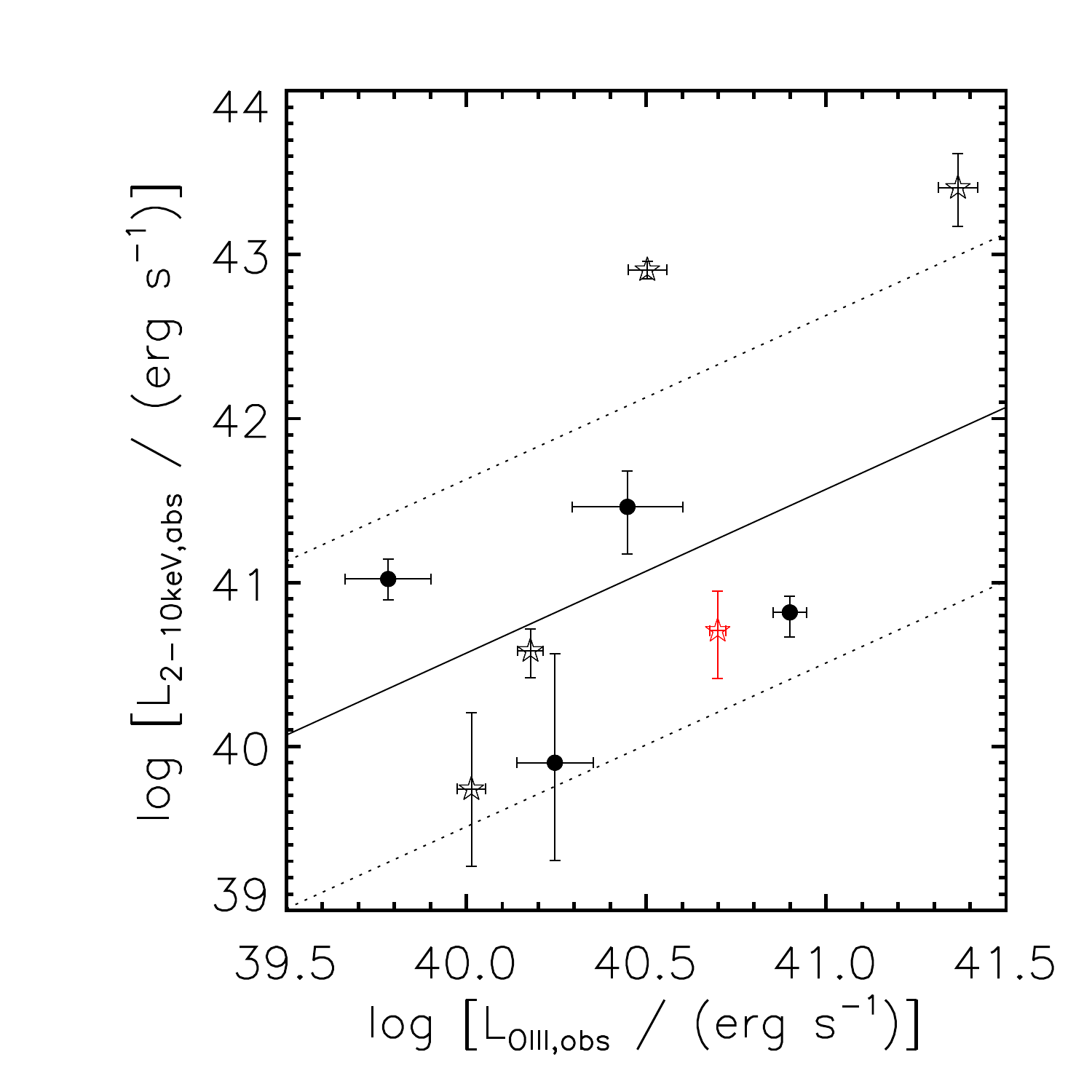}
\end{center}
\caption{Absorbed hard X-ray luminosity (2-10 kev) vs. observed \oiiiw luminosity for the galaxies with a central $3^{\prime\prime}$ fiber spectrum classified as star forming (open stars), and for the galaxies with a central $3^{\prime\prime}$ fiber spectrum classified as LINER (filled circles).  SDSS J1518+4244 NW, which is in a galaxy merger, is shown as the red star.  The other merger in our sample, SDSS J1407+4428, is not shown because we did not detect an X-ray source coincident with the location of its central $3^{\prime\prime}$ fiber.  The observed relation for optically-selected Type 2 AGNs is shown as the solid line, and the dotted lines illustrate the rms scatter in that relation \citep{HE05.1}.}  
\label{fig:Lx_Loiii}
\end{figure}

As Figure~\ref{fig:Lx_Loiii} shows, we find that our sample lies within the rms scatter of the fiducial relation for Type 2 AGNs, with the exceptions of two galaxies: SDSS J1630+2612 and SDSS J1430+5242.  For these two galaxies, the relatively high hard X-ray luminosities and low \oiii luminosities could be due to AGN flickering.  In this scenario, the spatially-extended AGN emission (traced by \oiiin) may be leftover from past AGN activity (e.g., \citealt{ER95.1,LI09.2}) while the central AGN emission (traced by hard X-rays) may be a sign of current AGN activity.  

We note that dust reddening in the host galaxy centers may also contribute, since the optical \oiii luminosity would be more affected by dust than the X-ray luminosity (since the X-ray sources are Compton thin; Table~\ref{tbl-4}).

\begin{figure}
\begin{center}
\includegraphics[width=3.in]{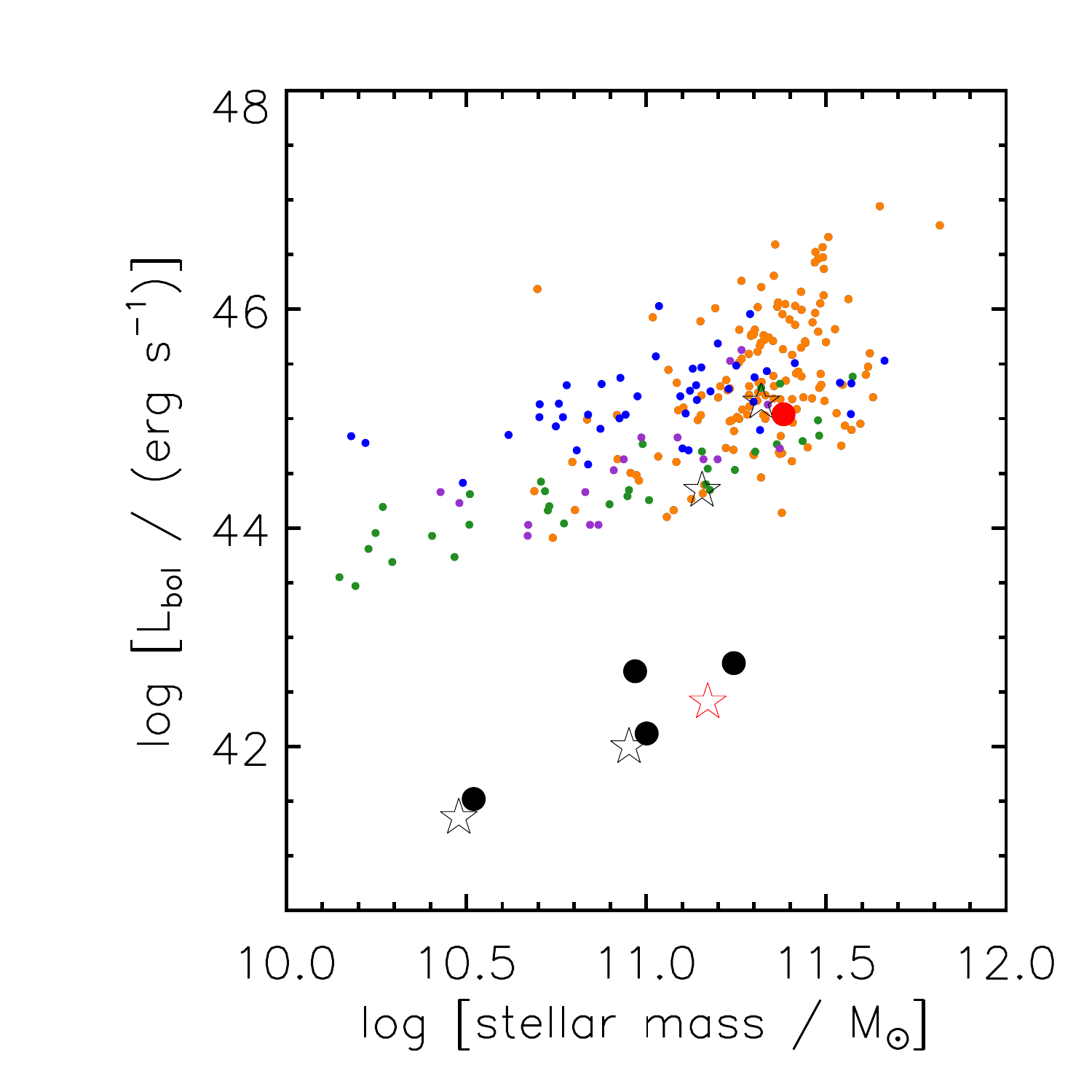}
\end{center}
\caption{AGN bolometric luminosity (derived from {\it Chandra} observations) vs. host galaxy stellar mass for the galaxies with a central $3^{\prime\prime}$ fiber spectrum classified as star forming (open stars), and for the galaxies with a central $3^{\prime\prime}$ fiber spectrum classified as LINER (filled circles).  Red symbols indicate the galaxy mergers, where for SDSS J1518+4244 (red open star) we have plotted the total bolometric luminosity from the two X-ray sources.  The small points show the AGNs in the MaNGA AGN catalog \citep{CO20.1}, which were identified via radio observations (orange), {\it WISE} colors (green), broad emission lines (blue), and Swift/BAT ultrahard X-ray detections (purple).  The bolometric luminosities of the small points come from bolometric corrections at a variety of wavelengths, depending on how each AGN was identifed (see \citealt{CO20.1} for details).}  
\label{fig:Lbol_stellarmass}
\end{figure}

\subsection{Galaxies with Off-nuclear Seyfert Regions Typically Host Low Luminosity AGNs} 

To understand whether the stellar masses or AGN bolometric luminosities of the off-nuclear Seyfert region galaxies are unusual, we compare them to the MaNGA AGN catalog, which is a catalog of MaNGA AGNs detected via {\it WISE}, {\it Swift}/BAT, radio, and/or broad emission lines \citep{CO20.1}.  The stellar mass ranges of the off-nuclear Seyfert region galaxies and the MaNGA AGN catalog galaxies are the same; the median log stellar mass is 11.1 for both samples.  According to a Kolmogorov-Smirnov test, there is a 82\% probability that the stellar masses of these two samples were drawn from the same distribution.

However, as Figure~\ref{fig:Lbol_stellarmass} shows, on average the AGN bolometric luminosities of the off-nuclear Seyfert region galaxies are significantly lower than the AGN bolometric luminosities of the MaNGA AGN catalog galaxies.  We conclude that the off-nuclear Seyfert regions are signatures of relatively low luminosity AGNs, where the AGN radiation in the central 3$^{\prime\prime}$ spectrum is not the dominant source of radiation (Section~\ref{nature}).

Two of the AGNs (SDSS J1407+4428 and SDSS J1630+2612) have $\log L_{bol} > 45$ and they fall in the same general bolometric luminosity and stellar mass parameter space as the MaNGA AGN catalog (Figure~\ref{fig:Lbol_stellarmass}); they are also {\it WISE}-detected AGNs  and we note that their {\it WISE}-derived bolometric luminosities (taken from \citealt{CO20.1}) are consistent with their {\it Chandra}-derived bolometric luminosities (Table~\ref{tbl-7}).  

There are two other AGNs (SDSS J0931+4904 and SDSS J1518+4244) that have {\it WISE}-derived bolometric luminosities that are two orders of magnitude higher than their {\it Chandra}-derived bolometric luminosities (Table~\ref{tbl-7}).  Since the {\it WISE}-derived bolometric luminosities are based on a mid-infrared to X-ray conversion that is not well constrained at the lower X-ray luminosities of these two AGNs \citep{ST15.1}, we suggest that the {\it Chandra}-derived bolometric luminosities are the more accurate luminosities for these two systems. Dust heated by star formation or by shocks (e.g., merger-induced shocks) can contribute to the mid-infrared emission (e.g., \citealt{PA17.1, BA21.1}), which may also partially explain the high {\it WISE}-derived bolometric luminosities for the star-forming galaxy SDSS J0931+4904 (SFR = 8 $M_\odot$ yr$^{-1}$; Table~\ref{tbl-1}) and the merger SDSS J1518+4244.

\subsection{Single-fiber BPT Diagnostics May Be Missing Half of AGNs}

Our parent sample is the 2727 galaxies that have been observed in MaNGA MPL-5.  From the single-fiber, central $3^{\prime\prime}$ spectrum alone, 130/2727 = 5\% of these galaxies are classified as hosting AGNs (Section~\ref{sample}).  However, another 173 galaxies were identified as AGN candidates based on spatially-resolved BPT and \ha analyses of their full maps of spaxels (for details see Section~\ref{sample}; \citealt{WY18.1}), even though their single-fiber central $3^{\prime\prime}$ spectrum did not show AGN signatures.  Here, we have studied a sample of 10 of these galaxies, and found that 70-100\% of them indeed host AGNs (Section~\ref{agn}).

If we assume that these results are representative of the full population of 173 AGN candidates, then another 121-173 galaxies (70-100\% of 173) in this sample may host AGNs.  In that case, the full AGN fraction would be (251-303)/2727$\sim10\%$.   With the caveat that our sample is small, our results suggest that single-fiber BPT diagnostics are potentially missing a significant number (up to a factor of 2) of AGNs in galaxies.

We find that there are multiple, often overlapping, reasons for single-fiber BPT diagnostics to fail to identify a complete population of AGNs.   First, the BPT classification is highly sensitive to the size and placement of the fiber: a fiber centered on a galaxy's nucleus may miss AGNs that are located outside the fiber diameter, and even in the case of a central AGN the BPT diagnostic can be dominated by emission from SF, post-AGB stars, shocks, and other ionization sources, depending on the fiber size and placement.  This domination of other (non-Seyfert) emission sources is heightened if the AGN is intrinsically low luminosity or dust obscured.  We also find evidence for AGN flickering in some of these galaxies, which is another avenue for other emission sources to dominate the BPT classification.  While the shut down is a brief phase in an AGN's lifetime (the light echo is visible for $\sim10^4-10^5$ years; \citealt{SC15.1}), it may be that selecting AGN candidates by their off-nuclear Seyfert signatures (as we have here) introduces a bias towards flickering AGNs.

\section{Nature of the 10 Galaxies}
\label{nature}

Here we interpret why each galaxy in our sample hosts an AGN, but has a central $3^{\prime\prime}$ spectrum that is classified as SF or LINER instead of Seyfert.  In summary, we find that in all five galaxies with central LINER spectra, the LINER-like emission line ratios are explained by post-ABG stars and/or shocks.  In one case (SDSS J1407+428), the off-nuclear Seyfert region is explained by an AGN in the nucleus of a companion galaxy.

The five galaxies with central SF spectra are a more complicated population.  In three of these we interpret SF as dominating because the AGN is flickering.  The other two galaxies with central SF spectra can be explained by SF emission dominating over a weak AGN, and SF emission dominating over an obscured AGN in a merger.

We developed these interpretations with the aid of images and emission line equivalent width and flux ratio maps of the 10 galaxies, which are shown in Figure~\ref{fig:maps}.

\subsection{SDSS J0743+4443} 

This galaxy hosts a weak AGN (Table~\ref{tbl-5}), and it also was not classified as an AGN by any other set of observations (Table~\ref{tbl-6}).
We interpret these results as evidence for a flickering AGN.  This explains why the X-ray source is weak, and why the central $3^{\prime\prime}$ spectrum shows evidence of SF but not AGN activity.  Further, the BPT map shows off-nuclear Seyfert emission outside of the plane of the host galaxy (Figure~\ref{fig:maps}), and this optical emission could be a light echo of past AGN activity.

\subsection{SDSS J0745+4604}

The LINER-like emission line ratios in this galaxy's central region are likely due to a combination of post-AGB stars and shocks, as the diagnostics in Figure~\ref{fig:maps} show.

\subsection{SDSS J0931+4904} 

This galaxy hosts a weak AGN (Table~\ref{tbl-5}) and has a relatively high SFR (8 $M_\odot$ yr$^{-1}$; Table~\ref{tbl-1}), which implies that the SF may be dominating over the AGN emission in the central region of the galaxy.  This would then explain the SF-like emission line ratios observed in the central $3^{\prime\prime}$ spectrum.

\subsection{SDSS J1254+4409}

This galaxy has evidence of shocks in its center (Figure~\ref{fig:maps}), which can explain why the central $3^{\prime\prime}$ spectrum is classified as LINER.

\subsection{SDSS J1407+4428} 

SDSS J1407+4428 is a major merger with two stellar bulges that have a projected separation of 8 kpc and a line-of-sight velocity difference of $\sim50$ km s$^{-1}$ (\citealt{EL17.1} measured $\sim30$ km s$^{-1}$, while \citealt{FU18.1} measured 76.8 km s$^{-1}$).

Using the MaNGA data, \cite{FU18.1} extracted a spectrum from a 2.6 kpc diameter circular aperture centered on each stellar bulge.  They classified the spectra using the \justniin-BPT diagnostic in combination with the WHAN diagram that relates \ha equivalent width and \justniin/\ha \citep{CI10.3}.  They found that the western nuclear spectrum is Seyfert, while the eastern nuclear spectrum is LINER.  We note that this result differs from the Portsmouth \justsiin-BPT classification that we use, which classifies the western nuclear spectrum as LINER.  \cite{FU18.1} classify this system as `apparent' binary AGNs, since the nuclear spectra could be explained by merger-driven shocks, a single AGN cross-ionizing its companion galaxy, or two separate AGNs.

\cite{EL17.1} also analyzed the MaNGA observations of this system and, via \justniin-BPT diagnostics (and supported by \justoin- and \justsiin-BPT diagnostics), found that the emission coincident with both of the stellar bulges is consistent with AGN photoionization.  This differs from the Portsmouth \justsiin-BPT classification of the western nuclear spectrum as LINER, perhaps because the Portsmouth classification uses a $3^{\prime\prime}$ fiber whereas the \cite{EL17.1} classification does not use an aperture.

In their analysis of the {\it Chandra}/ACIS observations, \cite{EL17.1} found two hard X-ray sources associated with the two stellar bulges.  For the eastern source, they measured $L_{2-10 \mathrm{keV}, \mathrm{unabs}}=(3.5 \pm 0.4) \times 10^{43}$ erg s$^{-1}$ and concluded that it is an AGN.  This is similar to our measurement of $L_{2-10 \mathrm{keV}, \mathrm{unabs}}=(5.5 \pm 1.1) \times 10^{43}$ erg s$^{-1}$, which we also used to conclude that the eastern X-ray source is an AGN.

For the western source, \cite{EL17.1} placed a $3^{\prime\prime}$ aperture on the position of the western galaxy bulge and detected 2 soft (0.3-8 keV) and 10 hard (2-8 keV) X-rays.  From this, they estimated a hardness ratio of 0.58 and $L_{2-10 \mathrm{keV}, \mathrm{unabs}}=(5.5 \pm 1.1) \times 10^{43}$ erg s$^{-1}$ (assuming a power-law spectrum with $\Gamma=1.8$).  Based on the optical emission line ratio diagnostics, hardness ratio, and X-ray luminosity, they concluded that the western X-ray source is likely an AGN.  Our search for a western X-ray source did not return a source detected at $>3\sigma$ above the background (Section~\ref{chandra}), so we cannot claim a detection of a western X-ray source.

In this system, the off-nuclear Seyfert region observed in the MaNGA data is explained by an AGN (the eastern X-ray source) in the nucleus of a companion galaxy, so that the Seyfert emission does indeed spatially coincide with an AGN.  The LINER-like emission in the western source is then explained by post-AGB stars and shocks (Figure~\ref{fig:maps}).

\subsection{SDSS J1430+5242} 

We interpret this galaxy as hosting a flickering AGN. Since the source was not classified as an AGN by any other set of observations (Table~\ref{tbl-6}), and the BPT map reveals spatially-extended Seyfert emission with a hole at the middle of the galaxy (Figure~\ref{fig:maps}), the spatially-extended emission could be an extended light echo of past AGN activity.  The X-ray luminosity is also higher than the canonical scaling relation between $L_{X,2-10\mathrm{keV}}$ and $L_{\oiiiwn}$ for AGNs (Section~\ref{LxLOIII}), which is evidence of new, current AGN activity.  The central bulge is obscured ($E(B-V)=0.78$; the second highest obscuration in our sample), and the SFR is relatively high (12 $M_\odot$ yr$^{-1}$; Table~\ref{tbl-1}), which would then explain why the central $3^{\prime\prime}$ spectrum is dominated by SF.

\subsection{SDSS J1518+4244} 

SDSS J1518+4244 is a major merger with two stellar bulges that have a projected separation of 5.5 kpc and a line-of-sight velocity difference of 91.9 km s$^{-1}$ \citep{FU18.1}.  Using the MaNGA data, \cite{FU18.1} extracted a spectrum from a 2.6 kpc diameter circular aperture centered on each stellar bulge and classified the spectra using the \justniin-BPT diagnostic along with the WHAN diagram \citep{CI10.3}.  They classified both nuclear spectra as starburst-AGN composite, which is different from the Portsmouth \justsiin-BPT classification of the northwestern nuclear spectrum as star-forming.   \cite{FU18.1} concluded that this system hosts `apparent' binary AGNs, where the nuclear spectra could be explained by merger-driven shocks, a single AGN cross-ionizing its companion galaxy, or two separate AGNs.

Here, we use {\it Chandra} observations to confirm that there are indeed dual AGNs in this galaxy merger (Table~\ref{tbl-5}).  The $3^{\prime\prime}$ SDSS fiber is centered on the northwest AGN, and that stellar bulge is obscured ($E(B-V)=0.81$; the highest obscuration in our sample), which helps explain why that fiber spectrum is dominated by SF-like emission line ratios.

\subsection{SDSS J1601+4521}

This galaxy has signs of shocks in the galaxy center (Figure~\ref{fig:maps}), which explain why the central $3^{\prime\prime}$ spectrum is classified as LINER.

\subsection{SDSS J1630+2612}

This galaxy's BPT map shows off-nuclear Seyfert emission outside of the plane of the host galaxy (Figure~\ref{fig:maps}), and the AGN has an X-ray luminosity that is higher than the canonical scaling relation between $L_{X,2-10\mathrm{keV}}$ and $L_{\oiiiwn}$ for AGNs (Section~\ref{LxLOIII}).  Consequently, this galaxy could host a flickering AGN that shut off in the past and left a light echo behind, and then turned on again and produced the X-ray emission. The SFR is relatively high (17 $M_\odot$ yr$^{-1}$; Table~\ref{tbl-1}), which would explain the SF classification of the central $3^{\prime\prime}$ spectrum.

\subsection{SDSS J1633+3911}

This galaxy hosts a weak AGN (Table~\ref{tbl-5}) and its central region has shock emission signatures (Figure~\ref{fig:maps}), which explain why the central $3^{\prime\prime}$ spectrum is classified as LINER.
 
\section{Conclusions}
\label{conclusions}

BPT diagnostics of a single-fiber spectrum of a galaxy are commonly used to identify AGNs.  Here, we have assembled a sample of 10 MaNGA galaxies where the \justsiin-BPT diagnostics of the central $3^{\prime\prime}$ fiber spectrum indicates SF or LINER, but not an AGN, while more than 10\% of the spaxels in the MaNGA observations are classified as Seyfert.  Thus, these galaxies are selected to have off-nuclear Seyfert regions but have central optical spectra that are not dominated by Seyfert emission.  We observed these 10 galaxies with {\it Chandra} to determine whether these off-nuclear Seyfert regions in fact indicate the presence of AGNs in these galaxies.

Our main results are summarized below.

1. The {\it Chandra} observations show that 7-10 (70-100\%) of the galaxies have confirmed AGNs, even though none of them were classified as AGNs based on \justsiin-BPT diagnostics of the single-fiber $3^{\prime\prime}$ spectrum.  While four of these galaxies also have mid-infrared {\it WISE} colors that indicate AGNs, the other six galaxies show no indication of AGNs via radio observations, broad emission lines, coronal lines, {\it WISE} colors, or {\it Swift}/BAT ultra hard X-ray detections (Table~\ref{tbl-6}).

2. These galaxies host AGNs with significantly lower bolometric luminosities, on average, than the bolometric luminosities of AGNs found in similar-mass MaNGA galaxies (Figure~\ref{fig:Lbol_stellarmass}).  The approach of selecting MaNGA galaxies with off-nuclear Seyfert regions seems to select for lower luminosity AGNs ($L_{bol} \sim 10^{42}$ erg s$^{-1}$). 

3. For the five galaxies that have their central $3^{\prime\prime}$ spectrum classified as LINER, the LINER line flux ratios are explained by emission from post-AGB stars and shocks.  Further, in one system, the off-nuclear Seyfert region of the galaxy spatially coincides with an AGN in the nucleus of a companion galaxy in an ongoing galaxy merger.

4. For the five galaxies that have their central $3^{\prime\prime}$ spectrum classified as SF, there are two reasons for the SF classification.  In three galaxies we find evidence for flickering AGNs, allowing SF to dominate the galaxy center while leaving remnant off-nuclear Seyfert emission in the AGN light echoes.  In the remaining two galaxies, the SF classification is due to central SF dominating over a weak or obscured AGN.

5.  Our results suggest that spatially-resolved spectroscopy may identify up to a factor of two more AGNs than single fiber spectra.  Due to the size and placement of the fiber, a single fiber spectrum centered on a galaxy's nucleus may miss AGNs in the nuclei of companion galaxies, low luminosity AGNs, dust obscured AGNs, and flickering AGNs.  In such cases, emission from star formation, post-AGB stars, and shocks can dominate the central regions of a galaxy, leading to a star forming or LINER classification of the central fiber spectrum. 

We conclude that single-fiber, BPT-based classifications of optical galaxy emission are missing large populations of AGNs, with biases towards missing AGNs in the nuclei of companion galaxies, low luminosity AGNs, obscured AGNs, and flickering AGNs.  Spatially resolved spectroscopy, provided in this case by MaNGA, shows that a significant portion of this missing populations of AGNs can be found via off-nuclear Seyfert emission regions. As more integral field spectrographs come online, it becomes ever more compelling and feasible to use spatially-resolved BPT diagnostics for a more complete census of AGNs.

\acknowledgements We thank the anonymous referee for comments that have improved the clarity of this paper.  Support for this work was provided by NASA through Chandra Award Number GO9-20089X issued by the Chandra X-ray Observatory Center, which is operated by the Smithsonian Astrophysical Observatory for and on behalf of NASA under contract NAS8-03060.  

The scientific results reported in this article are based in part on observations made by the Chandra X-ray Observatory, and this research has made use of software provided by the Chandra X-ray Center in the application packages CIAO, ChIPS, and Sherpa.  

Funding for the Sloan Digital Sky Survey IV has been provided by the Alfred P. Sloan Foundation, the U.S. Department of Energy Office of Science, and the Participating Institutions. SDSS-IV acknowledges support and resources from the Center for High-Performance Computing at the University of Utah. The SDSS web site is www.sdss.org.

SDSS-IV is managed by the Astrophysical Research Consortium for the  Participating Institutions of the SDSS Collaboration including the Brazilian Participation Group, the Carnegie Institution for Science, Carnegie Mellon University, the Chilean Participation Group, the French Participation Group, Harvard-Smithsonian Center for Astrophysics, Instituto de Astrof\'isica de Canarias, The Johns Hopkins University, Kavli Institute for the Physics and Mathematics of the Universe (IPMU) / University of Tokyo, the Korean Participation Group, Lawrence Berkeley National Laboratory, Leibniz Institut f\"ur Astrophysik Potsdam (AIP),  Max-Planck-Institut f\"ur Astronomie (MPIA Heidelberg), Max-Planck-Institut f\"ur Astrophysik (MPA Garching), Max-Planck-Institut f\"ur Extraterrestrische Physik (MPE), National Astronomical Observatories of China, New Mexico State University, New York University, University of Notre Dame, Observat\'ario Nacional / MCTI, The Ohio State University, Pennsylvania State University, Shanghai Astronomical Observatory, United Kingdom Participation Group, Universidad Nacional Aut\'onoma de M\'exico, University of Arizona, University of Colorado Boulder, University of Oxford, University of Portsmouth, University of Utah, University of Virginia, University of Washington, University of Wisconsin, Vanderbilt University, and Yale University.  This project makes use of the MaNGA-Pipe3D dataproducts. We thank the IA-UNAM MaNGA team for creating this catalogue, and the Conacyt Project CB-285080 for supporting them.

{\it Facilities:} \facility{{\it CXO}}

\appendix

\section{MaNGA Emission Line Maps and Spectra}

\begin{figure}
\centering
\includegraphics[width=6.5in]{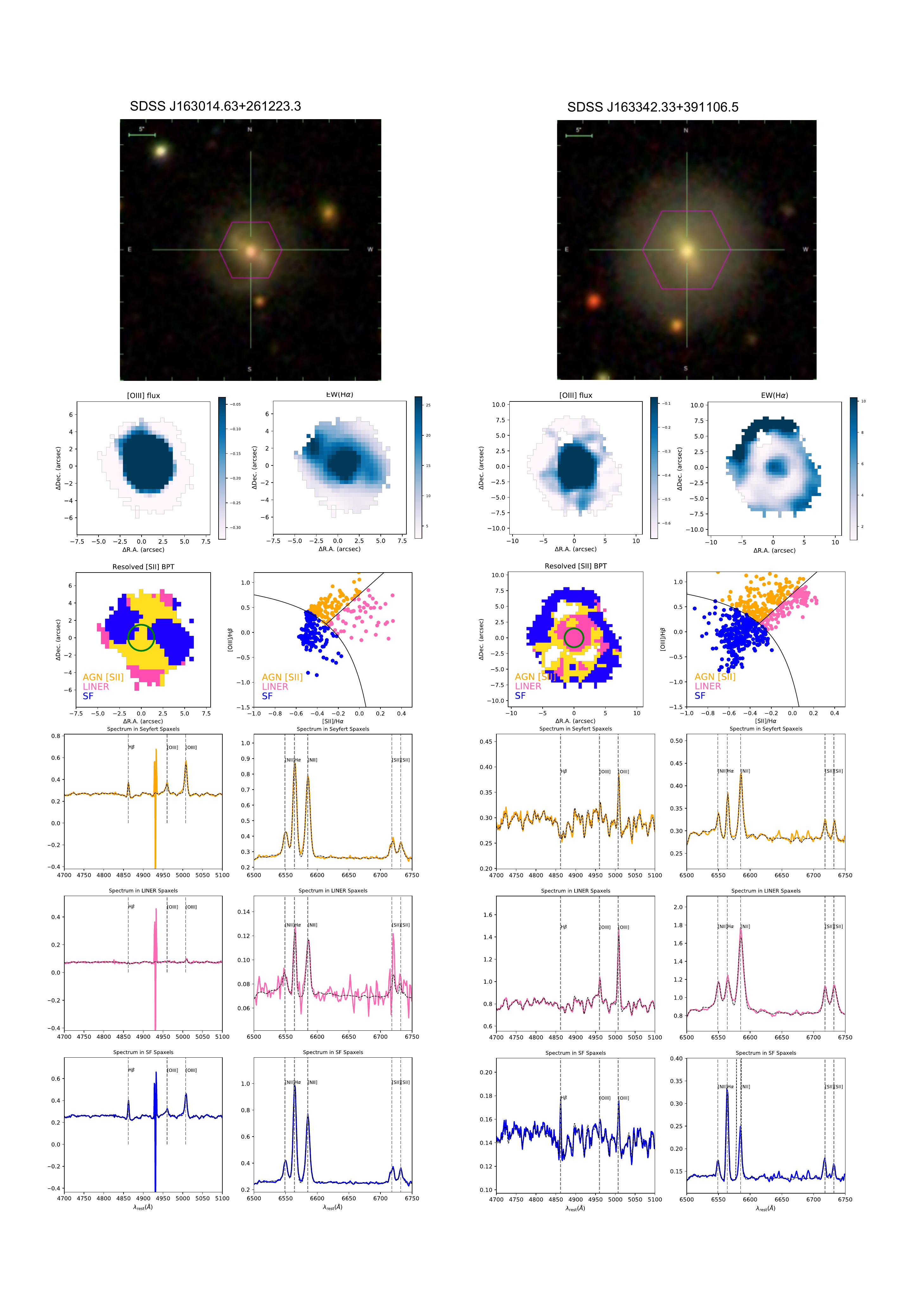}
\vspace{-0.5in}
\caption{MaNGA observations of the 10 galaxies.  Top: image of the galaxy with the MaNGA footprint shown as the purple hexagon.  Second row: \oiii flux map (left) and map of \ha equivalent width (right).  Third row: \justsiin-BPT diagram (left), where the green circle illustrates the $3^{\prime\prime}$ SDSS fiber, the Seyfert regions are shown in yellow, the LINER regions are shown in pink, and the SF regions are shown in blue.  Location of the Seyfert (yellow), LINER (pink), and SF (blue) spaxels on the \justsiin-BPT diagram (right).  Fourth row: the averaged spectrum of the Seyfert spaxels, shown over the wavelengths covering the \hb and \oiii lines (left) and the wavelengths covering the \justniin, \han, and \justsii lines (right).  Fifth and sixth rows: same as the fourth row, but for the LINER spaxels and the SF spaxels, respectively.}
\label{fig:manga_maps}
\end{figure}

The properties of the emission lines observed in the MaNGA spaxels are a key part of our analysis of these 10 galaxies.  Here, we show spectra and \justsiin-BPT diagrams for each galaxy in Figure~\ref{fig:manga_maps}.

\begin{figure}
\begin{center}
\includegraphics[width=6.5in]{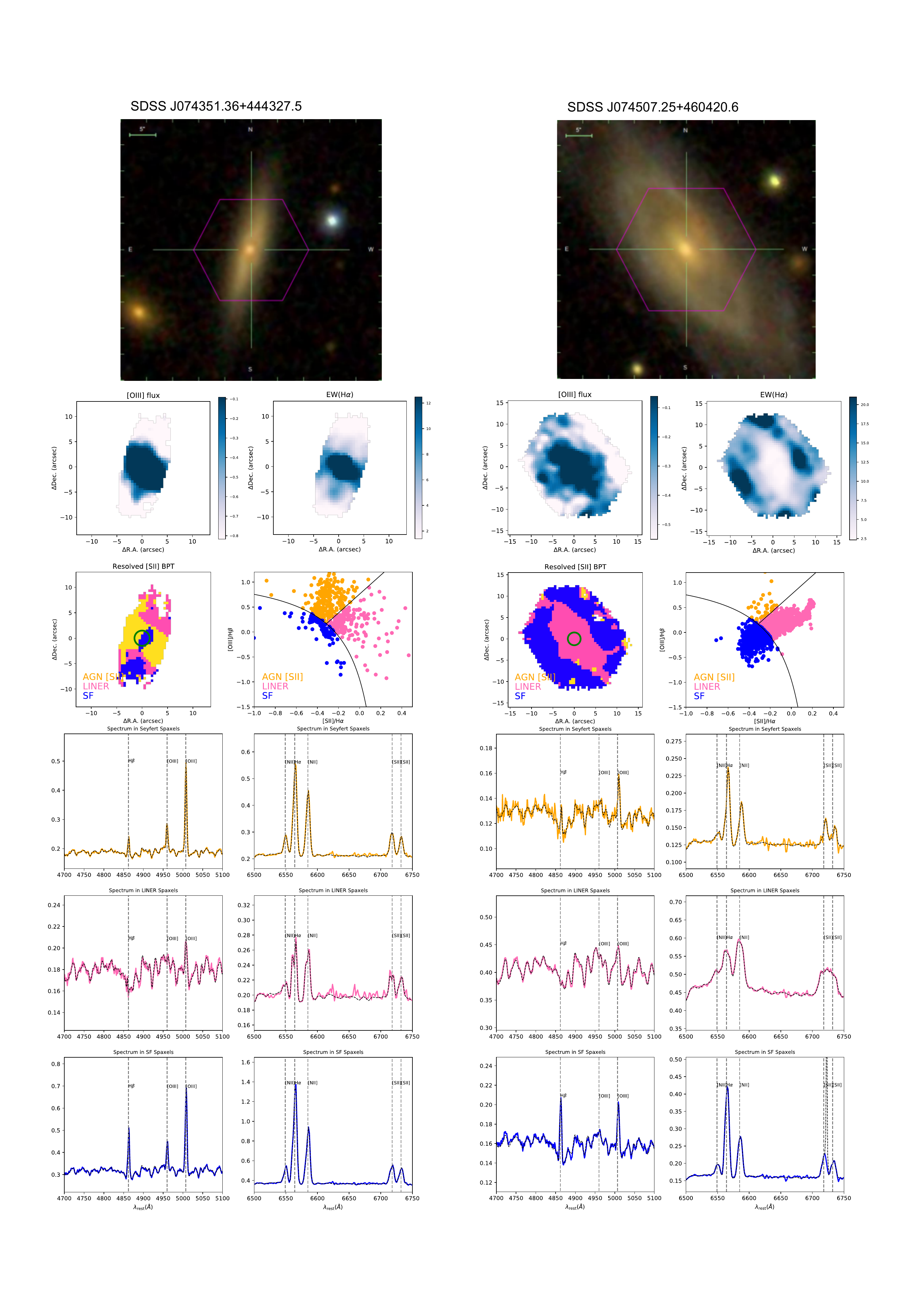}
\end{center}
\end{figure}

\begin{figure}
\centering
\includegraphics[width=6.5in]{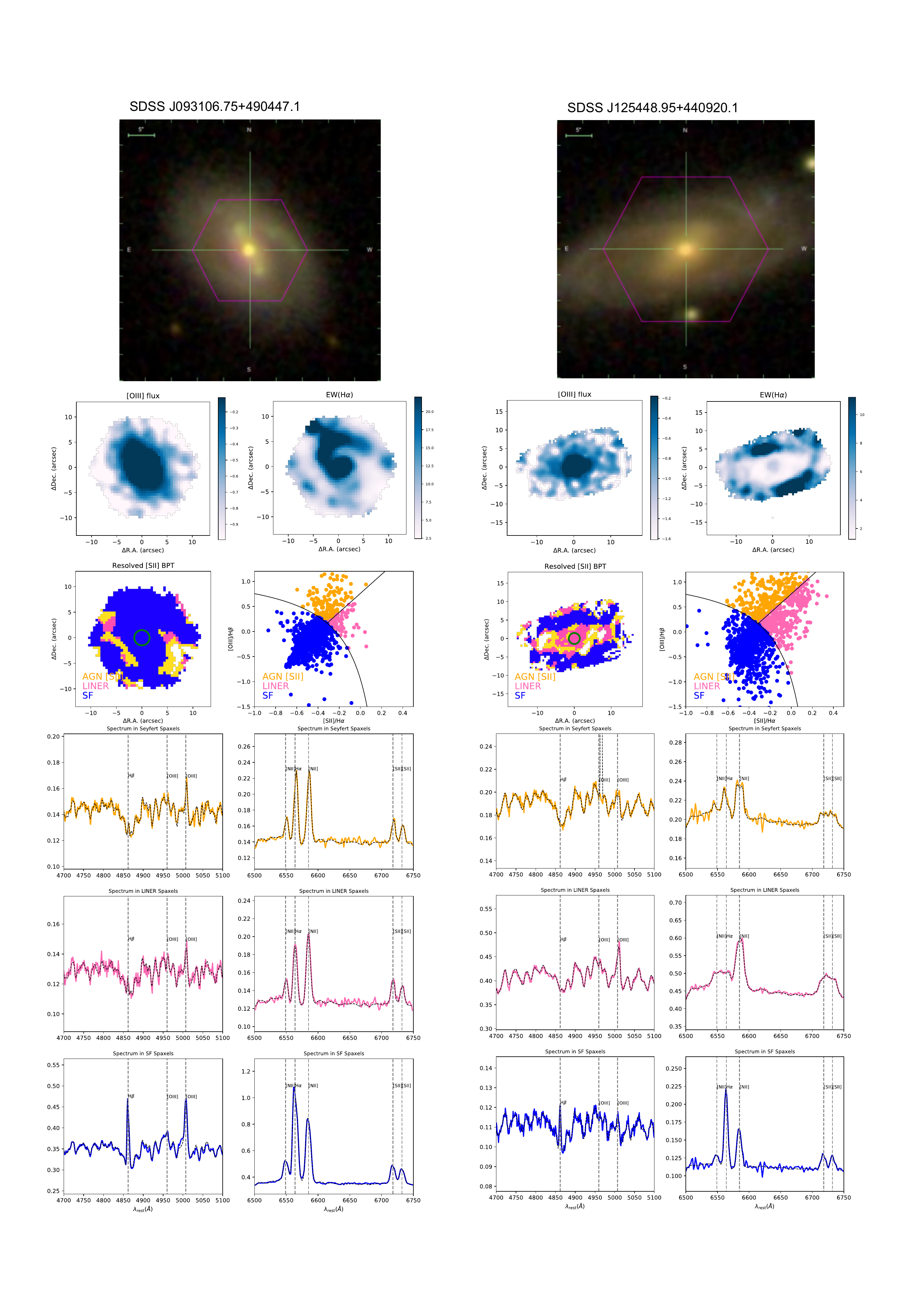}
\end{figure}

\begin{figure}
\centering
\includegraphics[width=6.5in]{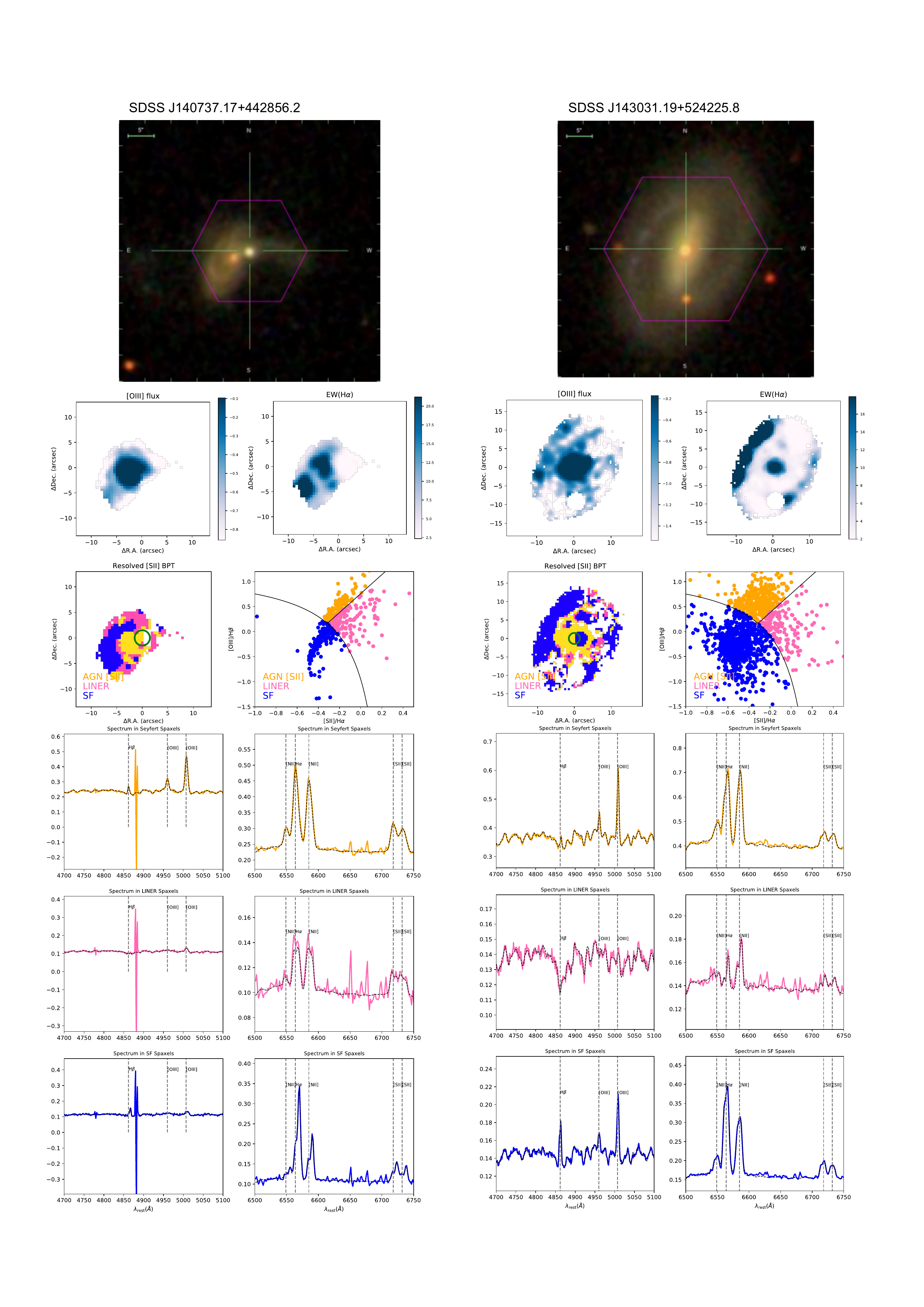}
\end{figure}

\begin{figure}
\centering
\includegraphics[width=6.5in]{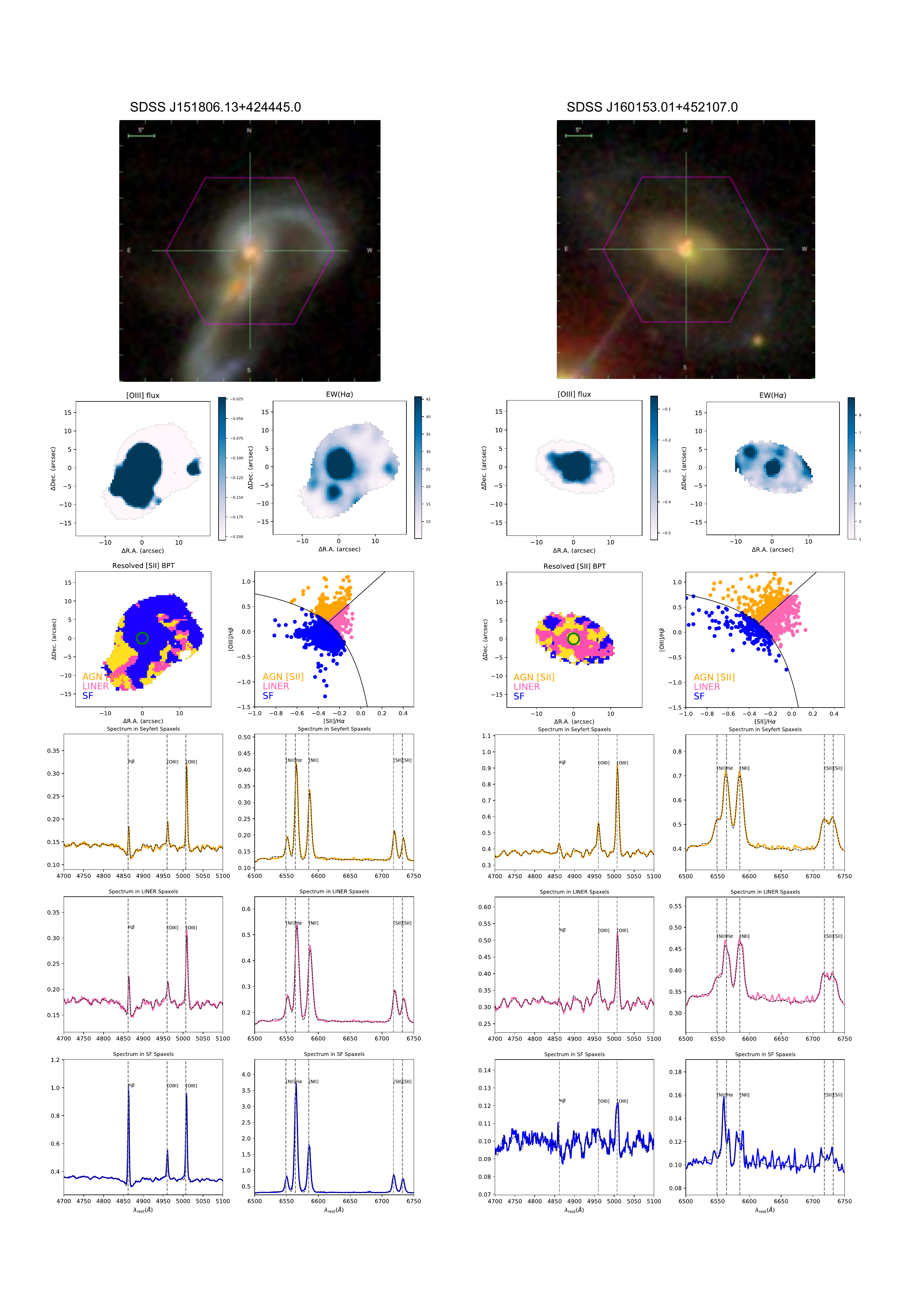}
\end{figure}

\section{{\it Chandra} Spectra}

\begin{figure}
\renewcommand{\thefigure}{5}
\begin{center}
\includegraphics[width=6.5in]{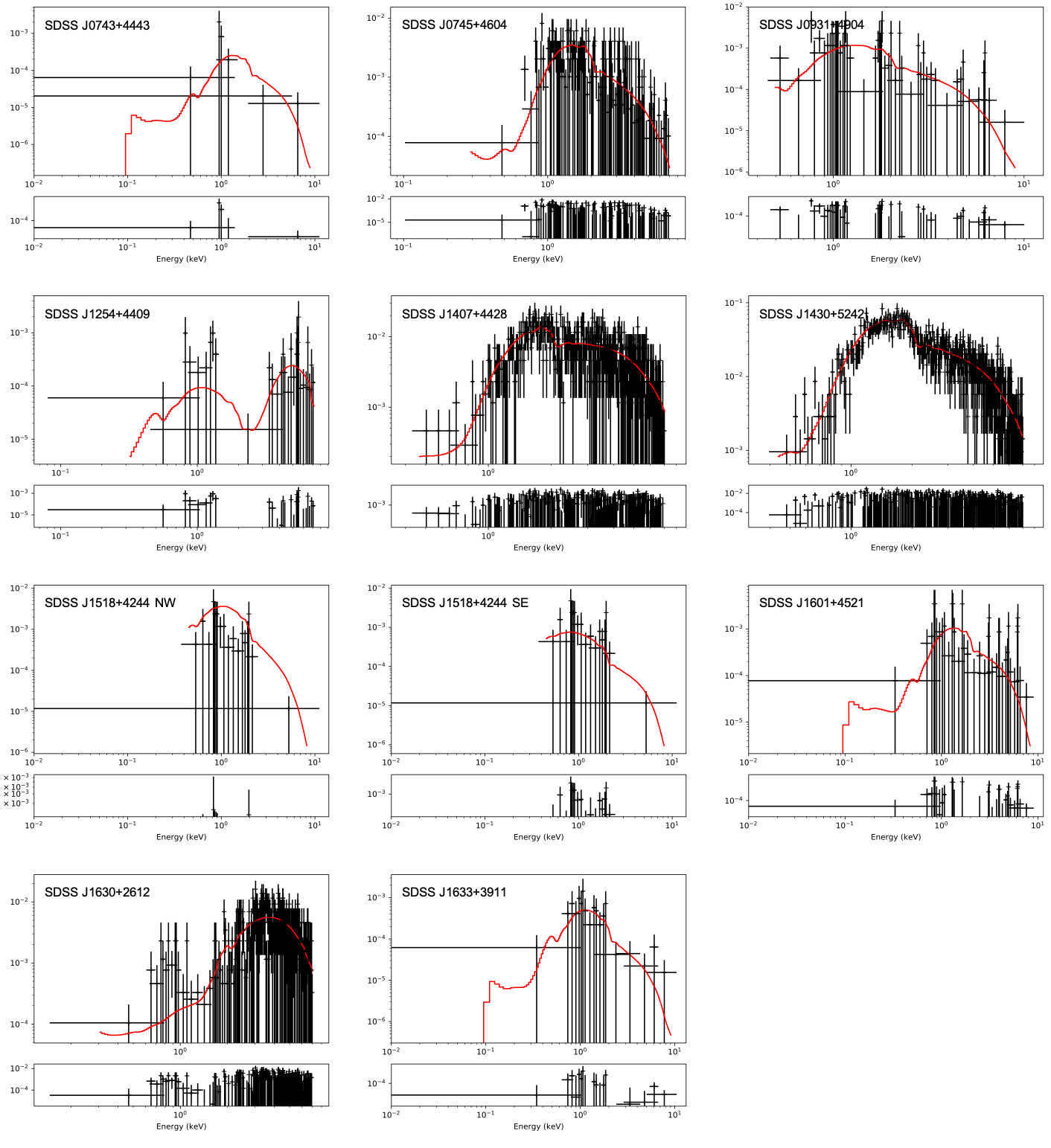}
\end{center}
\caption{{\it Chandra} X-ray spectra of the 10 galaxies, including separate spectra for the NW and SE sources in SDSS J1518+4244.  The red lines show the best fits to the spectra, with best-fit parameters given in Table~\ref{tbl-4}.}
\label{fig:chandra_spectra}
\end{figure}

The best fits to the {\it Chandra} spectra for the 10 galaxies are shown in Figure~\ref{fig:chandra_spectra}.

\bibliographystyle{aasjournal}

\end{document}